\documentclass[]{aa}
\usepackage{hyperref}
\usepackage{natbib}
\usepackage{graphicx}
\usepackage{epsfig}
\usepackage{color}
\usepackage{txfonts}
\usepackage{rotating}

\begin{document}
\newcommand{\micron}{$\mu$m}
\newcommand{\HII}{H~{\sc ii}}
\newcommand{\Gnaught}{G$_0$}
\newcommand{\cii}{[C~{\sc ii}]}
\newcommand{\oi}{[O~{\sc i}]}
\newcommand{\ci}{[C~{\sc i}]}

\titlerunning{Herschel spectroscopy of Galactic PDRs}
\authorrunning{D. J. Stock et al.}

\title{{\em Herschel}\thanks{{\it Herschel} is an ESA space observatory with science instruments provided by European-led Principal Investigator consortia and with important participation from NASA.} PACS and SPIRE spectroscopy of the Photodissociation Regions associated with S~106 and IRAS~23133+6050}

\author{D.~J.~Stock\inst{1}, M.~G.~Wolfire\inst{2}, E.~Peeters\inst{1}\fnmsep\inst{3}, A.~G.~G.~M.~Tielens\inst{4}, B.~Vandenbussche\inst{5} , C.~Boersma\inst{6} \& J.~Cami\inst{1}\fnmsep\inst{3}}

\institute{Department of Physics and Astronomy, University of Western Ontario, London, ON, N6A 3K7, Canada; \email{dstock4@uwo.ca}
\and
Department of Astronomy, University of Maryland, College Park, MD 20742, USA        
\and
SETI Institute, 189 Bernardo Avenue, Suite 100, Mountain View, CA 94043, USA
\and
Leiden Observatory, Leiden University, P.O. Box 9513, NL-2300 RA, The Netherlands
\and
Instituut voor Sterrenkunde, Katholieke Universiteit Leuven, Celestijnenlaan 200D, 3001 Leuven, Belgium
\and
Space Science Division, MS 245-6, NASA Ames Research Center, Moffett Field, CA 94035, USA	
}

\date{Version 1.2: Accepted 23 April 2015}

\abstract
   {Photodissociation regions (PDRs) contain a large fraction of all of the interstellar matter in galaxies. Classical examples include the boundaries between ionized regions and molecular clouds in regions of massive star formation, marking the point where all of the photons energetic enough to ionize hydrogen have been absorbed.}
   {To determine the physical properties of the PDRs associated with the star forming regions IRAS~23133+6050 and S~106 and present them in the context of other Galactic  PDRs associated with massive star forming regions.}
   {We employ Herschel PACS and SPIRE spectroscopic observations to construct a full 55--650 \micron\ spectrum of each object from which we measure the PDR cooling lines, other fine- structure lines, CO lines and the total far-infrared flux. These measurements (and combinations thereof) are then compared to standard PDR models. Subsequently detailed numerical PDR models are compared to these predictions, yielding additional insights into the dominant thermal processes in the PDRs and their structures. }
   {We find that the PDRs of each object are very similar, and can be characterized by a two-phase PDR model with a very dense, highly UV irradiated phase ($n \sim$ 10$^6$ cm$^{-3}$, \Gnaught~$\sim$ 10$^{5}$) interspersed within a lower density, weaker radiation field phase ($n \sim$ 10$^4$ cm$^{-3}$, \Gnaught~$\sim$ 10$^{4}$). We employed two different numerical models to investigate the data, firstly we used RADEX models to fit the peak of the $^{12}$CO ladder, which in conjunction with the properties derived yielded a temperature of around 300 K. Subsequent numerical modeling with a full PDR model revealed that the dense phase has a filling factor of around 0.6 in both objects. The shape of the $^{12}$CO ladder was consistent with these components with heating dominated by grain photoelectric heating. An extra excitation component for the hightest J lines (J > 20) is required for S~106. }
   {}   

   {}

   \maketitle


\section{Introduction}

Understanding photodissociation regions (PDRs; also known as photon dominated regions; e.g. \citealt{1985ApJ...291..722T, 1989ApJ...338..197S}) is an essential part of comprehending the physics and chemistry of the interstellar medium (ISM). The classical definition of a PDR is the region in which the energy balance is dominated by far-UV photons with energies less than that required to ionize hydrogen (13.6 eV). In practice this definition means that classical PDRs form the `skin' of the ionized \HII\ regions and represent the buffer zone between the fully ionized \HII\ region material and molecular clouds, typically this means that they occur at A$_\textrm{V}$ $<$ 5 with respect to the cloud surface.  In a broader sense, a large fraction of all neutral ISM material exists under these conditions, underlining the importance of understanding the physics and chemistry of these regions.  

The prototypical PDR, the Orion (Bar) region, has been extensively studied (e.g., \citealt{1985ApJ...291..747T, 1993ApJ...404..219S, 1995A&A...294..792H, 1995A&A...303..541J,1997A&A...327L...9S, 2011MNRAS.410.1320R, 2012ApJ...753..168B, 2013A&A...550A..57O, 2014arXiv1405.5553A, 2014arXiv1405.3903N}); and reviews of the whole complex by \citealt{1989ARA&A..27...41G, 2001ARA&A..39...99O, 1993Sci...262...86T, 2010A&A...518L.116H}). This has been largely due to two factors: a) its close proximity (450 pc, \citealt{1989ARA&A..27...41G}) and b) the fact that the Orion Bar is viewed almost exactly edge on. This combination allows for the relatively easy separation of the emission from each region across the transition from ionized to PDR material and subsequent comparison with detailed PDR models (e.g., \citealt{1993Sci...262...86T}). These models are highly sophisticated codes that include a detailed chemical network as well as descriptions of relevant heating and cooling processes. As a function of the incident radiation field strength (usually denoted \Gnaught\footnotemark) and density (or thermal pressure) the PDR models calculate the thermal equilibrium temperature and dominant atomic and molecular species in chemical balance. The models also predict the dominant emission line strengths that can be compared with observations. From this data the PDR models synthetic spectra are constructed and compared to those observed. It is now generally agreed that most PDRs are clumpy, that is, they take the form of dense clumps (n > 10$^5$ cm$^{-3}$) embedded within an interclump medium 10-100 times less dense \citep{1990ApJ...365..620B, 1993ApJ...405..216M, 2006A&A...451..917R}. Such two phase models can reproduce many of the observed atomic and molecular emission from PDRs.

\footnotetext{\Gnaught\ is defined in units of the Habing field after \citet{1968BAN....19..421H}, in which the local interstellar UV radiation field integrated between 6 and 13.6 eV was found to have an average strength of 1.2 $\times$ 10$^{-4}$ ergs cm$^{-2}$ s$^{-1}$ sr$^{-1}$. The mean interstellar radiation field $\chi$ \citep{1978ApJS...36..595D}, is greater than the Habing field by a factor of 1.7. }

Measurements of PDR diagnostics have been sparse in previous decades because features diagnostic of PDR physical conditions predominantly emit in the far-infrared (FIR) spectral regime at wavelengths between $\sim$30 and 1000 \micron. This wavelength region is notoriously difficult to observe, as much of this range requires above atmosphere observations, using space-based observatories such as ISO, Spitzer and Herschel, balloons such as BICE \citep{1998ApJS..115..259N} or aircraft like the KAO (Kuiper Airborne Observatory) and SOFIA (Stratospheric Observatory for Infrared Astronomy) to cover the shortest wavelengths and ground based sub-mm observatories to cover the longer wavelength emissions. In recent years this has become much easier though, as the Herschel space observatory (\citealt{2010A&A...518L...1P}) was able to collect full spectra in the PDR-significant range. 

In this paper we expand the sample of PDRs observed by Herschel, using the PACS and SPIRE instruments (\citealt{2010A&A2...518L...2P, 2010A&A2...518L...3G}), to include two Galactic PDRs generated by massive star formation (IRAS~23133+6050 and S~106) in order to understand the physical conditions and compare their properties to those of other Galactic PDRs, including the prototypical Orion Bar. We obtained spectra of the important atomic PDR cooling lines ([O~{\sc i}] 63, 145 \micron, [C~{\sc ii}] 158 \micron, and [C~{\sc i}] 609 \micron) as well as the CO and $^{13}$CO ladders from mid- to high-J. These observations allow us to probe various depths into the respective PDRs, with the atomic lines providing constrains on the warm, dense surface layers at low A$_\textrm{V}$ (e.g. T = 100-1000 K, n = 10$^4$--10$^5$ cm$^{-3}$, A$_\textrm{V}$ $<$ 4) and the CO ladders probing both the warm, low A$_V$ regions, as well as the deeper, cooler regions at higher A$_\textrm{V}$ (e.g. T = 10-100 K, n = 10$^4$--10$^5$ cm$^{-3}$, A$_\textrm{V}$ $>$ 4 ). The nature of our targets and prior knowledge of them is outlined in Section~\ref{sec:targs}. A description of the observations, data reduction procedures and the measurement of the spectra is given in Section~\ref{sec:dra}. The results of this work, starting with the PDR diagnostics including the molecular tracers such as $^{12}$CO are discussed in Section~\ref{sec:res} in terms of the physical conditions of these PDRs. Subsequently we compare these observations to the outputs of a sophisticated PDR model in Section~\ref{sec:num}. Discussion of these results is presented in terms of the geometry of the PDRs and the derived conditions with those of other Galactic PDRs in Section~\ref{sec:disc}. Finally, we present our conclusions and a short summary in Section~\ref{sec:concsum}.



\section{IRAS~23133+6050 and S~106}\label{sec:targs}

\begin{figure*}
  \centering
    \includegraphics[width=18cm]{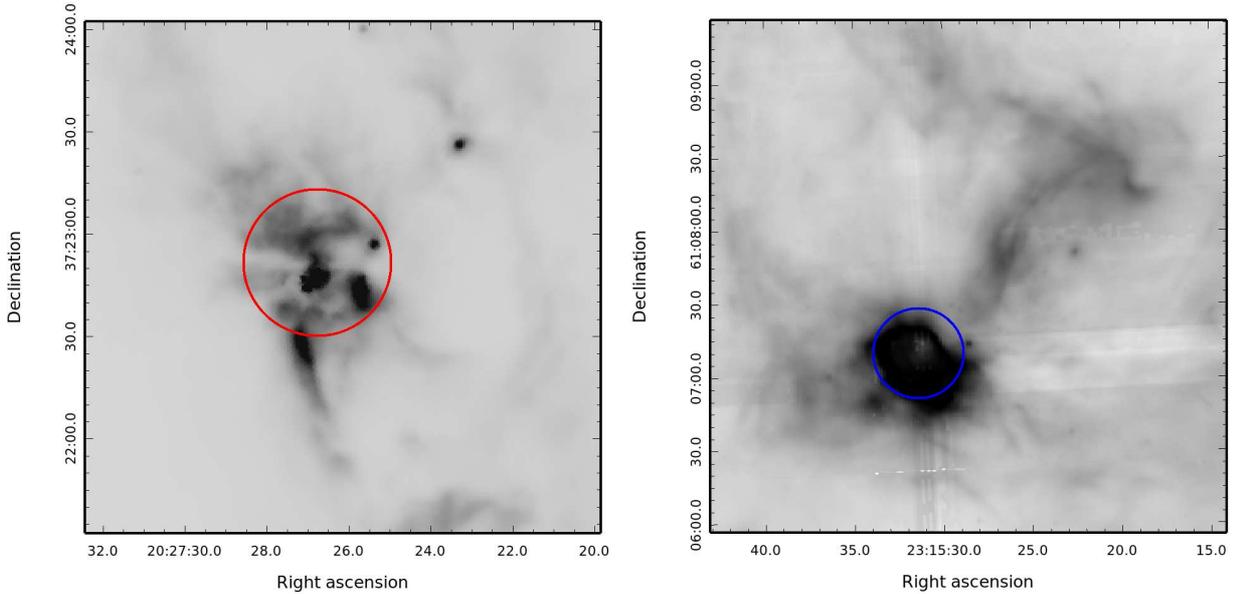}
    \caption{Pointing for S~106 (left) and IRAS~23133+6050 (right). Overlaid on each GLIMPSE \citep{2003PASP..115..953B, 2009PASP..121..213C} IRAC [8.0] image is the corrected SPIRE beam aperture (43\arcsec\ and 37\arcsec\ respectively). We adopt the convention of coloring points relevant to S~106 and IRAS~23133+6070 red and blue respectively, beginning with the apertures in this Figure. } 
    \label{fig:pointings}
\end{figure*}

Both IRAS~23133+6050 and S~106 are well-studied regions of star formation. In this section we review the pertinent literature in order to guide our later discussion. The properties drawn from the literature are summarized in Table~\ref{tables:sources}. GLIMPSE survey images \citep{2003PASP..115..953B,2009PASP..121..213C}, obtained using the Spitzer/IRAC instrument \citep{2004ApJS..154...10F}, of each source are shown in Fig.~\ref{fig:pointings}.

\begin{table*}
\caption{\label{tables:sources}Source Characteristics}
\begin{center}
\begin{tabular}{p{5.5cm}cccc}
\hline\hline
                          & S~106  & Reference & IRAS~23133+6050 & Reference \\
\hline
\textit{Astrometric:} & & & & \\
Right Ascension [h:m:s]                           & 20:27:26.8                              &                & 23:15:33.1          &   \\
Declination [$^\circ$: \arcmin: \arcsec]          & +37:22:49                               &                & +61:07:18           &   \\
Angular Size of \HII\ region [\arcsec]            & $\sim$ 120 $\times$ 60                  & 1              & 9                   & 4 \\
Distance [pc]                                     & 1500                                    & 2, 3           & 5500                & 4 \\
Physical Radius of \HII\          region [pc]     & 0.44$^a$                                & 2              & 0.1                 & 4 \\
Physical Radius of Herschel Beam [pc]             & 0.15                                    &                & 0.49                &   \\
& & & & \\
\textit{Central Stars:}& & & & \\
Name                                              & S106-IR                                 & 5              & IRAS~23133+6050     & \\
Spectral Type                                     & O8                                      & 6              & O8.5                & 9\\
                                                  & B0 - O8                                 & 7              & O9.5                & 10\\
                                                  & O7 - O9                                 & 8              & O8.5                & 11\\
& & & & \\
\textit{ISM Properties:}& & & & \\
Electron density [cm$^{-3}$]                      & 1.5 -- 6 $\times$ 10$^3$                & 12             & 2 $\times$ 10$^3$   & 2\\
                                                  &                                         &                & 8.6 $\times$ 10$^3$ & 13\\
PDR density   [cm$^{-3}$]                         & 1 -- 3.5 $\times$ 10$^5$                & 1              &                     &       \\
                                                  & 0.3 -- 5 $\times$ 10$^6$                & 14             &                     & \\
                                                  & $\sim$ 10$^5$                           & 4              &                     & \\
\hline
\end{tabular}
\end{center}
\bigskip
$^a$: Here we have adopted 60\arcsec\ as an average.
\tablebib{(1) \citet{2012A&A...542L..12S}; (2) \citet{2005AA...436..569N}; (3) \citet{2007AA...474..873S} ; (4) \citet{2003AA...407..957M}; (5) \citet{1993MNRAS.262..839R}; (6) \citet{2000AA...358.1035V}; (7) \citet{1979A&A....74...89E}; (8) \citet{1982ApJ...254..550G}; (9) \citet{1999ApJ...514..232K}; (10) \citet{1994ApJS...91..659K}; (11) \citet{2002ApJS..138...35H}; (12) \citet{1978AA....63..325I}; (13) \citet{1994ApJS...91..659K}; (14) \citet{2003AA...406..915S}.}

\end{table*}

\subsection{IRAS~23133+6050}
IRAS~23133+6050\footnotemark, appears to be a classical example of an isolated star forming region within a molecular cloud (e.g., \citealt{1994ApJS...91..659K, 1999ApJ...514..232K} show it to be an isolated ultra-compact \HII\ region). From an astrometric perspective, there has been some debate regarding the distance of IRAS~23133+6050. Early estimates by \citet{1978ApJS...38..309H} gave the distance to the Cassiopeia OB2 star forming region, of which IRAS~23133+6050 is part, at 2.63 kpc. \citet{2002A&A...381..571P} found a distance of 5.5 kpc using kinematical methods and ISO-SWS spectroscopy, which agrees with an earlier calculation by \citet{1984ApJ...279..125F} who found d = 5.2 $\pm$ 0.9 kpc. However other authors, e.g., \citet{2001ApJ...560..806L}, adopt a smaller value of 3.1 kpc, closer to initial estimates, based on spectrophotmetric observations of associated stars as derived by \citet{1993AA...275...67B}. We adopt the higher value throughout this paper.

\footnotetext{We note that while the designation IRAS~23133+6050 technically refers to the exciting star, throughout this paper we will also refer to its \HII\ region / PDR complex by this name as appears to be the common practice in the literature. The \HII\ region can be found in SIMBAD as `[KCW94]~111.612+0.374' reflecting the \citet{1994ApJS...91..659K} radio detection of the region.}

\citet{1999ApJ...514..232K} showed that the radio emission from IRAS~23133+6050 is consistent with the \HII\ region being powered by a late O star of spectral type O9.5. This star powers a cometary ultra compact \HII\ region with a diameter of 0.2 pc \citep{2003AA...407..957M} based on their calculated distance of 5.5 kpc. Using ISO observations of the MIR [S~{\sc iii}] lines \citet{2003AA...407..957M} found the electron density of the \HII\ region to be about 2 $\times$ 10$^3$ cm$^{-3}$. Comparatively few studies have discussed the PDR / molecular cloud envelope of the IRAS~23133+6050 region. Assuming pressure equilibrium, the expected density contrast between a typical \HII\ region and its PDR is expected to be around 30; which leads to an expected PDR density of the order of 6 $\times$ 10$^4$ cm$^{-3}$. IRAS~23133+6050 was included in a sample of UC-\HII\ regions observed in the sub-mm by SCUBA, providing fluxes at 450 and 850 \micron\ \citep{2006A&A...453.1003T}.


\subsection{S~106}

S~106 (Sh 2-106; \citealt{1959ApJS....4..257S}) is an extremely complex source, consisting of multiple components and IR sources. S~106 has been found to be at a distance of 1.2--1.8 kpc \citep{2007AA...474..873S}, we adopt 1.5 kpc following \citet{2005AA...436..569N}. S~106 is believed to be powered mainly by a single O-type star (S106-IR; \citealt{1993MNRAS.262..839R}), which is surrounded by a very small optically thick disk \citep{1996ASPC...93...47H}. The double lobed structure of the \HII\ region is created by the shadows of this disk. 

The PDR and molecular cloud components of S~106 have been the focus of prior studies (see \citealt{2008hsf1.book...90H} for a thorough review). The overall structure of the S~106 system is very similar to bipolar nebulae formed by a combination of stellar outflows and a circumstellar disk. That the central star is undergoing significant mass loss has long been known (e.g., \citealt{1993MNRAS.265...12D,2007MNRAS.380..246G, 2009MNRAS.400..629P}) which is consistent with the idea that the large ionized lobes were sculpted in this way. However the origin of the central dark lane has been more controversial. 

Early models of the S~106 system invoked a large optically thick torus of dense material as the explanation for the double lobed appearance of the system and central dark lane. Higher resolution radio/sub-mm observations (e.g., \citealt{1993MNRAS.262..839R}) dispelled this notion though, and replaced it with the idea that the central regions of S~106 are dominated by a variety of clumps of denser material, which have in turn been suggested to be debris from the dissolution of a large dense torus \citep{1995MNRAS.277..307L}. In subsequent years it has been suggested that the dark lane is the projected shadow of an edge on accretion disk\footnotemark\ around the central young star (e.g., \citealt{2005AA...436..569N}), however recent SOFIA results seem to be inconsistent with this model \citep{2012A&A...542L..12S}. 

\footnotetext{Other authors have detected the signature of such a disk without commenting on the issue of the projected shadow / dark band structure, e.g., \citet{2013MNRAS.436..511M}.}

Subsequently several authors have investigated the PDR environment of S~106 each using different datasets (e.g., \citealt{2000AA...358.1035V,2003AA...406..915S,2005AA...436..569N}). The first of these, by \citet{2000AA...358.1035V}, used large aperture ISO-SWS and LWS (\citealt{1996A&A...315L..49D,1996A&A...315L..38C}) to find the global properties of the FIR emission from S~106. They conclude, on the basis of very strong [Si~{\sc ii}] emission, that the PDR associated with S~106 must have densities of around 10$^5$ cm$^{-3}$ and UV radiation field strength of greater than 10$^5$. \citet{2000AA...358.1035V} also conclude that this PDR must be 5-15\arcsec\ away from the central star. A subsequent study by \citet{2003AA...406..915S} showed that the PDR of S~106 is much more extended than previously thought, extending up to several arcminutes from the central star. In addition, \citet{2003AA...406..915S} used PDR diagnostics to show that a two phase PDR model was necessary to understand the line emission from S~106. They concluded that a very dense molecular component with an H$_2$ density greater than 3 $\times$ 10$^5$ cm$^{-3}$ was necessary, along with a less dense interclump phase. This finding agreed with the previous study of CO emission by the same authors which found evidence of a very dense component \citep{2002AA...384..225S}. Parallel efforts to those using higher spatial resolution mid-IR instruments have yielded further insights as to the central structures. \citet{2005AA...436..569N} studied S~106 using near-infrared spectro-imaging of a combination of hydrogen recombination lines and molecular hydrogen emission. From the perspective of this work, the \citet{2005AA...436..569N} study came to two key results: 1) the observations of H$_{2}$ lines were found to be consistent with shocks driven by outflows of the central star and, 2) the internal structure of the S~106 region based on their line measurements (presented in their Fig. 12). Shocks would be an expected part of any high-mass star forming region, however previous studies found little evidence for their presence (e.g., \citealt{2000AA...358.1035V}).



\section{Observations, Data Reduction and Analysis}\label{sec:dra}

\subsection{Observations}

A log of all Herschel observations used is presented in Table~\ref{tables:obs}. The observations were performed as part of Herschel program \verb+OT1_epeeters_01+. The objects were chosen to provide a sample to detect the long wavelength PAH features. This paper concerns the PDR properties of those objects which were not included in other PDR surveys. 

The SPIRE observations each used the pointed spectroscopy instrument mode with no mapping to fill in the spatial gaps between the pixels. The PACS observations used the spectroscopy `Range Scan' mode, which provides a complete spectrum from 50--200 \micron\ without any spectral gaps. These observations were combined to produce a complete spectrum for each object covering from 50--650 \micron\ for the area on the sky covered by the central SPIRE spaxel beam after correcting for constant beam size. The pointings for each object are shown in Fig.~\ref{fig:pointings} along with the corrected SPIRE beam size for each source (see Section~\ref{sec:drspire}). The spectral resolution varied as a function of wavelength for each instrument, with SPIRE providing resolutions between $\sim$400 and $\sim$1300 and PACS giving resolutions between 1000 and 5000. In practice this means that none of the detected lines were spectrally resolved.

\begin{table*}
\caption{\label{tables:obs}Observation Log}
\begin{center}
\begin{tabular}{c c c c c c c c}
\hline\hline
Object & Date & Instrument & Mode& Obs. ID & Duration (s) &$\lambda$ Range (\micron) & Comment \\
\hline
\\
S~106       & 2011-05-25 & SPIRE & Spectro Point & 1342221683 & 528 & 190--650 \\ 
S~106       & 2011-11-20 & PACS  & Range Scan    & 1342232575 & 670 & 50--74  \& 100--145 \\
S~106       & 2011-11-20 & PACS  & Range Scan    & 1342232576 & 339 & 50--74  \& 100--145 & background\\
S~106       & 2011-11-20 & PACS  & Range Scan    & 1342232577 & 669 & 70--105 \& 140--210 & background \\
S~106       & 2011-11-20 & PACS  & Range Scan    & 1342232578 & 1330 & 70--105 \& 140--210 \\
\\
IRAS 23133 & 2011-08-15 & SPIRE & Spectro Point & 1342227453 & 1068 & 190--650 \\ 
IRAS 23133 & 2012-01-14 & PACS  & Range Scan & 1342237480 & 670 & 50--74  \& 100--145 \\
IRAS 23133 & 2012-01-14 & PACS  & Range Scan & 1342237481 & 339 & 50--74  \& 100--145 & background\\
IRAS 23133 & 2012-01-14 & PACS  & Range Scan & 1342237482 & 669 & 70--105 \& 140--210 & background \\
IRAS 23133 & 2012-01-14 & PACS  & Range Scan & 1342237483 & 1330 & 70--105 \& 140--210 \\
\hline
\end{tabular}

\end{center}

\end{table*}

\subsection{Data Reduction}


\paragraph{Spire}\label{sec:drspire}

The SPIRE data were reduced using the Herschel Interactive Processing Environment (HIPE; version 11.0; \citealt{2010ASPC..434..139O}). Each of our observations was taken in the `Bright Source' mode, which enables SPIRE to observe sources which would usually saturate the detectors. The reduction process used the ``SPIRE Spectrometer Single Pointing User Reprocessing'' HIPE script. This allows the user to calibrate bright source mode observations and use the newest version of the SPIRE calibration data which at the time of reduction was \verb+SPIRE_CAL_11_0+.  

Accurate calibration of our sources was achieved by using the HIPE semi-extended correction tool which implements the methodology described by \citet{2013A&A...556A.116W}. The semi-extended correction tool calculates the degree to which the observed source differs from a point source using the break between the orders of the SPIRE detectors where the beam size changes. In order to use this tool it was necessary to reduce the observations using the `point source' calibration rather than the `extended' calibration. The semi-extended correction tool results in a spectrum which has been corrected to a constant beam size, 43\arcsec\ for S~106 and 37\arcsec\ for IRAS~23133+6050. This eliminates the flux jump between the two orders of the SPIRE instrument and allows easier cross calibration with the PACS data. The semi-extended correction tool only corrects the fluxes for the central pixel as the beams of the long and short wavelength detectors are centered with respect to each other. For each of the surrounding, off-axis, SPIRE pixels for both objects also contain semi-extended structure (e.g., Fig.~\ref{fig:pointings}) and as such they cannot be reliably calibrated at this time due to this software limitation.

\paragraph{PACS}

 The PACS data were reduced using HIPE version 11.0 and the PACS calibration set \verb+PACS_CAL_56_0+. The observations of both S~106 and IRAS~23133+6050 were performed using the unchopped spectroscopy mode with corresponding background observations (see Table~\ref{tables:obs}) because both sources are extended beyond the maximum PACS chop throw. Exposure times and ObsIDs are given in Table~\ref{tables:obs}.

The PACS data are affected by several systematics, mainly order leakage and spectral ghosts\footnotemark. The regions of order leakage were trimmed as necessary. This principally affected the longer wavelength spectra in band R1 (105--210 \micron), where it proved impossible to have a well calibrated spectrum beyond around 190 \micron. Regions affected by these problems were trimmed from the final data, resulting in gaps in the spectral coverage.

\footnotetext{see the PACS data processing guide: \\ \url{http://herschel.esac.esa.int/Data\_Processing.shtml}}

It was found that the continuum was offset slightly in each PACS spectral order, likely a reflection of the varying beam size, and as such the orders were scaled such that they agreed with the longest wavelength order. It was found that the offsets between the orders were additive, in that the line fluxes for lines in overlapping regions matched before scaling. For S~106, the blue orders agreed with the final red order (B2A, B2B, R1B. Hence the only scaling necessary was for R1A, which had to be increased by the addition of 850 Jy to agree with the continua of the other orders. For the IRAS~23133+6050 measurements, the first PACS spectral order was found to have a higher background than expected resulting in a strong offset of 950 Jy with respect to the continuum of the other bands. While these values sound large, they represent factors of only around 20\%.

The final PACS spectra were created by coadding the pixels to match size of the beam produced by the SPIRE semi extended source correction. This was accomplished by calculating the relative contributions of each pixel to this beam and summing all of the pixels in the correct proportions. The final fluxes were achieved by scaling this coadded PACS spectrum such that the continuum slopes and intensities matched. This scaling factor was found to be multiplicative in order for the fluxes of the $^{12}$CO lines to be consistent across the PACS/SPIRE gap. For S~106 this factor was relatively small (1.5), while for IRAS~23133+6050 it was much greater (4). 


\subsection{Analysis}

\begin{figure*}
	\centering
	\includegraphics[width=17cm]{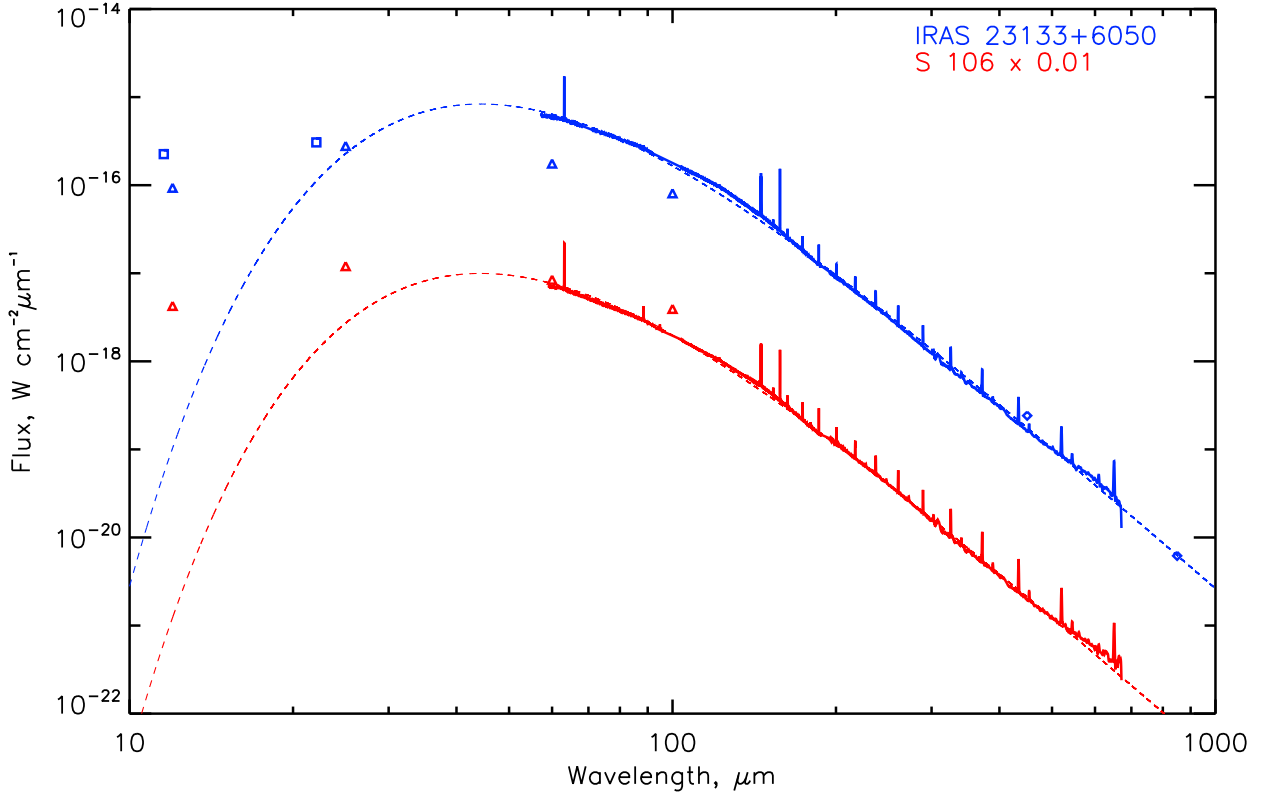}
	\caption{Full PACS + SPIRE spectra for IRAS~23133+6050 (top, blue) and S~106 (bottom, red, multiplied by 0.01 for clarity). In addition we include the IRAS photometric points for each source (triangles), and for IRAS~23133+6050 also the WISE photometry \citep{2010AJ....140.1868W} as squares and the SCUBA observations of \citet{2006A&A...453.1003T} as diamonds. A modified blackbody fit for each spectrum is shown as a dashed line. }
    \label{fig:seds}
\end{figure*}

\begin{table*}
\caption{\label{tables:fluxes}PDR Line and Continuum Flux Measurements}
\begin{center}
\begin{tabular}{l c c c c}
\hline\hline
Species & Transition$^a$ & Wavelength  & \multicolumn{2}{c}{ Flux }\\
        &            & [\micron]   & \multicolumn{2}{c}{[10$^{-5}$ erg s$^{-1}$ cm$^{-2}$ sr$^{-1}$]} \\
        &            &            & IRAS~23133+6050 & S~106 \\
\hline
     $^{12}$CO  &     4 -- 3 &   650.25 &   5.67 $\pm$   0.43 &   6.96 $\pm$   0.16 \\
     $^{12}$CO  &     5 -- 4 &   520.23 &   8.11 $\pm$   0.18 &   10.6 $\pm$   0.1 \\
     $^{12}$CO  &     6 -- 5 &   433.56 &   11.6 $\pm$   0.1  &   14.3 $\pm$   0.1 \\
     $^{12}$CO  &     7 -- 6 &   371.65 &   16.4 $\pm$   0.2  &   20.1 $\pm$   0.3 \\
     $^{12}$CO  &     8 -- 7 &   325.23 &   20.6 $\pm$   0.6  &   26.6 $\pm$   0.5 \\
     $^{12}$CO  &     9 -- 8 &   289.12 &   28.1 $\pm$   0.3  &   29.6 $\pm$   0.5 \\
     $^{12}$CO  &    10 -- 9 &   260.24 &   35.6 $\pm$   0.5  &   39.2 $\pm$   0.5 \\
     $^{12}$CO  &   11 -- 10 &   236.61 &   37.7 $\pm$   0.6  &   40.4 $\pm$   1.0 \\
     $^{12}$CO  &   12 -- 11 &   216.93 &   45.0 $\pm$   1.0  &   50.5 $\pm$   1.7 \\
     $^{12}$CO  &   13 -- 12 &   200.27 &   45.7 $\pm$   2.5  &   54.1 $\pm$   3.4 \\
     $^{12}$CO  &   14 -- 13 &   186.00 &   47.2 $\pm$   0.8  &   57.3 $\pm$   0.5 \\
     $^{12}$CO  &   15 -- 14 &   173.63 &   42.4 $\pm$   0.7  &   49.5 $\pm$   0.4 \\
     $^{12}$CO  &   16 -- 15 &   162.81 &   35.8 $\pm$   1.4  &   44.5 $\pm$   2.4 \\
     $^{12}$CO  &   17 -- 16 &   153.27 &   28.9 $\pm$   1.0  &   35.9 $\pm$   0.9 \\
     $^{12}$CO  &   18 -- 17 &   144.78 &   20.6 $\pm$   2.5  &   24.1 $\pm$   1.7 \\
     $^{12}$CO  &   19 -- 18 &   137.20 &   17.7 $\pm$   1.6  &   19.1 $\pm$   0.4 \\
     $^{12}$CO  &   20 -- 19 &   130.37 &   12.9 $\pm$   1.0  &   13.6 $\pm$   0.4 \\
     $^{12}$CO  &   21 -- 20 &   124.19 &   6.43 $\pm$   2.00 &   11.7 $\pm$   0.5 \\
     $^{12}$CO  &   22 -- 21 &   118.58 &   4.11 $\pm$   1.53 &   10.3 $\pm$   0.8 \\
     $^{12}$CO  &   23 -- 22 &   113.46 &   4.34 $\pm$   2.97 &   8.74 $\pm$   0.89 \\
\\
     $^{13}$CO  &     5 -- 4 &   544.16 &   1.83 $\pm$   0.13 &   1.81 $\pm$   0.21 \\
     $^{13}$CO  &     6 -- 5 &   453.50 &   2.49 $\pm$   0.13 &   2.91 $\pm$   0.06 \\
     $^{13}$CO  &     7 -- 6 &   388.74 &   2.14 $\pm$   0.18 &   2.47 $\pm$   0.14 \\
     $^{13}$CO  &     8 -- 7 &   340.18 &   3.66 $\pm$   0.42 &   5.52 $\pm$   0.49 \\
     $^{13}$CO  &     9 -- 8 &   302.41 &   5.53 $\pm$   1.57 &   5.37 $\pm$   1.38 \\
     $^{13}$CO  &    10 -- 9 &   272.20 &   4.59 $\pm$   0.45 &   6.77 $\pm$   0.91 \\
     $^{13}$CO  &   11 -- 10 &   247.49 &   5.29 $\pm$   0.63 &   6.05 $\pm$   0.83 \\
     $^{13}$CO  &   12 -- 11 &   226.90 &   3.47 $\pm$   0.88 &   4.45 $\pm$   1.10 \\
\\
   ~[C~{\sc i}] &              $^3$P$_1$ -- $^3$P$_0$ &   609.14 &     1.36 $\pm$   0.09 &     0.86 $\pm$   0.14 \\
   ~[C~{\sc i}] &              $^3$P$_2$ -- $^3$P$_1$ &   369.87 &     3.09 $\pm$   0.23 &     2.72 $\pm$   0.27 \\
  ~[C~{\sc ii}] &  $^2$P$_{^3/_2}$ -- $^2$P$_{^1/_2}$ &   157.74 &    604.7 $\pm$   3.5  &    391.1 $\pm$   2.0 \\
   ~[O~{\sc i}] &              $^3$P$_1$ -- $^3$P$_0$ &   145.54 &    439.6 $\pm$   2.1  &    447.9 $\pm$   1.7 \\
  ~[N~{\sc ii}] &              $^3$P$_2$ -- $^3$P$_1$ &   121.91 &     34.3 $\pm$   2.1  &     34.7 $\pm$   1.3 \\
 ~[O~{\sc iii}] &              $^3$P$_1$ -- $^3$P$_0$ &    88.36 &     34.2 $\pm$   4.6  &    233.1 $\pm$   2.3 \\
   ~[O~{\sc i}] &              $^3$P$_2$ -- $^3$P$_1$ &    63.18 &     2934 $\pm$   10   &   2987   $\pm$   7 \\
\\
$\Sigma$ FIR &  --  & 10 -- 1000 & 1.86 $\times$ 10$^{6}$ & 1.69 $\times$ 10$^{6}$\\
\hline
\end{tabular}
\end{center}

\bigskip
$^a$: Upper and lower J values for $^{12}$CO and $^{13}$CO.

\end{table*}

\subsubsection{Line Fluxes}\label{sec:line_flux}
In each of the coadded final spectra, the spectral lines were analyzed in the following ways. The line fluxes were extracted via fitting Gaussians and local continua to the data. It was found that better fits resulted from fitting a linear local continuum to regions of the spectrum on either side of the line, and then fixing the linear parameters fitted for these regions when fitting the full spectrum of a Gaussian line profile plus linear continuum. The signal to noise estimates for each line measurement were calculated by using the root mean square (rms) noise in the continuum spectral windows after the subtraction of the linear fit. This rms value can then be combined with the integrated line flux, width and the instrumental resolution to yield the final signal to noise ratio for the measured line. There is only one significant blended pair of lines present in the data, the 371.65 \micron\ CO 7-6 and 370.42 \micron\ [C~{\sc i}] 2-1 lines. Initially we performed a fit using two gaussians to the complex. We have checked this fit by calculating the flux of the 7-6 CO line using the neighboring CO lines and subtracting this flux from the total flux of the blended complex. The residual of this process yielded the same \ci\ flux as the initial fitting procedure. The measured line fluxes are given in Table~\ref{tables:fluxes}.

For S~106 we can compare the measured PDR cooling line fluxes here to those found by other observations in the past. The most direct comparison can be drawn to those presented by \citet{2000AA...358.1035V} and \citet{2003AA...406..915S} which are based on observations by ISO-LWS, KAO-FIFI and KOSMA-SMART (\citealt{1996A&A...315L..38C, 1991IJIMW..12..859P, 2002stt..conf..143G}). A direct comparison between the measured surface brightnesses is given in Table~\ref{tables:flux_comp}. Some care should be taken in comparing the presented quantities as the beam sizes differ by large factors from the Herschel measurements presented here. For the dominant cooling line, [O~{\sc i}] 63 \micron, it is easy to see why the ISO-LWS surface brightness measurements would be much smaller as they were measured using a larger beam by a factor of around five, and hence the overall surface brightness is lower as the flux of S~106 is peaked strongly in the central regions near the exciting source IRS4. \citet{2000AA...358.1035V} measured surface brightnesses for a separate set of ISO-LWS observations which were also centered on S~106 and found that the [O~{\sc i}] 63 \micron\ line had a surface brightness between the values found by \citet{2003AA...406..915S} and our measurements. Evidently different pointings around the S~106 region can result in very different results, even considering the fact that the comparatively large LWS beam encompasses the majority of strong PDR cooling line emission from the S~106 complex (e.g. see the Sofia/GREAT map of [C~{\sc ii}] 158 \micron\ presented by \citet{2012A&A...542L..12S}).

Overall, our observations of the [O~{\sc i}] line agree within $\sim$30\% with the LWS measurements if we assume that the S~106 flux is concentrated at the center of the beam and compensate for the different beam sizes. The same process yields similar results for the other LWS cooling line measurements -- for example for the [C~{\sc ii}] 158 \micron\ line our results agree with the LWS observations almost exactly for the \citet{2000AA...358.1035V} pointing and to within 30\% for the \citet{2003AA...406..915S} pointing, assuming that the beam size quoted by \citet{2003AA...406..915S} is correct (see footnote to Table~\ref{tables:flux_comp}). Overall then, we find agreement within the quoted absolute flux uncertainties of 30\% for both ISO-LWS and PACS (LWS: \citealt{1996A&A...315L..43S}; PACS: PACS Spectrometer Calibration Document\footnotemark). At first glance, the KAO surface brightness measurement of [O~{\sc i}] (from \citealt{2003AA...406..915S}) that we include in Table~\ref{tables:flux_comp} seems incompatible with our measurement, as it was for a beam smaller than the SPIRE beam by a factor of three and found a \textit{lower} surface brightness. However, \citet{2003AA...406..915S} discuss this problem as the discrepancy between KAO/Herschel is almost the same as between the KAO and LWS measurements. \citet{2003AA...406..915S} conclude that the KAO and LWS measurements can be reconciled by taking into account the spatially varying emission intensities -- specifically that if one assumed a gaussian profile for the [O~{\sc i}] emission the LWS beam size, the LWS flux could be recovered from the KAO measurement.

\footnotetext{http://herschel.esac.esa.int/twiki/pub/Public/PacsCalibrationWeb/ PacsSpectroscopyPerformanceAndCalibration\_v2\_4.pdf}

\begin{table*}
\caption{\label{tables:flux_comp} Prior Measurements of PDR cooling lines in S~106}
\begin{center}
\begin{tabular}{l c c c c c c}
\hline\hline
Line & Wavelength & Herschel Brightness                         & Herschel Beam  & Prior Measurement                            & Beam           & Instrument   \\
     & [\micron]  & [10$^{-5}$ erg s$^{-1}$ cm$^{-2}$ sr$^{-1}$] & [10$^{-8}$ sr] & [10$^{-5}$ erg s$^{-1}$ cm$^{-2}$ sr$^{-1}$] & [10$^{-8}$ sr]  \\
\hline
\\
\multicolumn{2}{l}{\textit{\citet{2000AA...358.1035V}:}}\\
~[O~{\sc i}]  & 63    &       2987 $\pm$   7  & 3.41  & 1125 $\pm$ 114    & 16.3$^a$ & ISO-LWS     \\
~[O~{\sc i}]  & 63    &       2987 $\pm$   7  & 3.41  & 1125 $\pm$ 114    & 3.41$^b$ & ISO-LWS     \\
~[O~{\sc i}]  & 146   &   447.9 $\pm$   1.7   & 3.41  & 288.8 $\pm$ 29.0  & 8.8$^a$  & ISO-LWS     \\
~[C~{\sc ii}] & 158   &   391.1 $\pm$   2.0   & 3.41  & 157.2 $\pm$ 15.9  & 11.6$^a$ & ISO-LWS     \\
\\
\multicolumn{2}{l}{\textit{\citet{2003AA...406..915S}:}}\\
~[O~{\sc i}]  & 63    &       2987 $\pm$   7  & 3.41  & 639 $\pm$ 7       & 12.4     & ISO-LWS     \\
~[O~{\sc i}]  & 63    &       2987 $\pm$   7  & 3.41  & 2100              & 1.3      & KAO-FIFI    \\
~[O~{\sc i}]  & 146   &    447.9 $\pm$   1.7  & 3.41  & 74 $\pm$ 3        & 9.0      & ISO-LWS     \\
~[C~{\sc ii}] & 158   &    391.1 $\pm$   2.0  & 3.41  & 80                & 8.6      & KAO-FIFI    \\
~[C~{\sc ii}] & 158   &    391.1 $\pm$   2.0  & 3.41  & 107 $\pm$ 3       & 8.5      & ISO-LWS     \\
~[C~{\sc i}]  & 609   &    0.86 $\pm$   0.14  & 3.41  & 0.217             & 7.1      & KOSMA       \\
~[C~{\sc i}]  & 370   &    2.72 $\pm$   0.27  & 3.41  & 2.64              & 7.1      & KOSMA       \\
\hline
\end{tabular}

\end{center}
\bigskip
$^a$: \citet{2003AA...406..915S} quote smaller values for the area of the LWS beam at the same wavelengths than \citet{2000AA...358.1035V}, which is likely due to the ISO-LWS beam profiles being significantly different in practice than expected, leading to changes to the quoted beam sizes in the LWS documentation over time. For example see the discussion presented by \citet[][Section 3.1]{2006A&A...446..561L}. As such, the values presented by \citet{2003AA...406..915S} are likely to be closer to the true LWS beam sizes.
\end{table*}

\subsubsection{Total FIR Flux}
The total flux of the far-infrared dust continuum was determined by fitting a modified blackbody to each spectrum. In order to better constrain the peak of this spectrum, we searched for additional observations of our sources in the 20-60 \micron\ regime such that we might acquire extra data points via aperture photometry using the SPIRE beam. However, realistic measurements of the fluxes for these sources were not available as they are very bright and saturated in both MIPS and WISE data. Figure~\ref{fig:seds} includes IRAS photometric points, however these are integrated totals for the entire source and not scaled to the Herschel beam. In addition we investigated using MSX photometry, however the MSX fluxes agree well with the IRAS points. We therefore investigated using spectral data from ISO-SWS and LWS to help constrain the shorter wavelength regime. The ISO spectrometers used very different beam sizes though, with the LWS aperture being of $\sim$ 120\arcsec\ diameter while the SWS aperture was 20\arcsec\ $\times$ 33\arcsec at its longest wavelengths. This corresponds to very different areas of each object dominating the beam. For SWS the bright central parts of the nebulae fill the beam, while for LWS the colder outer areas fill the beam. For S~106, the LWS spectra agree well with the final PACS spectrum, albeit with an offset of around  50\%. This is likely because there is extended bright structure in the LWS beam and as such it gives a similar spectral slope. For IRAS~23133+6050, the picture is less clear, as the PACS spectrum has a much steeper gradient than the LWS spectrum. IRAS~23133+6050 is much more compact than S~106, with all of the hot material being confined to the center. The shallower gradient of the LWS continuum then may reflect the extra cold material in the beam.  

For each object the peak of the SED falls just before the PACS wavelength range, therefore in order to measure the total infrared flux we fit a modified blackbody to the spectrum and integrated all of the flux of the fitted blackbody between 10 and 1000 \micron. The aperture mismatches with the existing ISO-SWS and -LWS data and our observations proved impossible to reconcile and as such we have no strong constraints on the blue end of the spectrum. As a test of this uncertainty, we performed an additional fit in which the FIR flux was fit with two blackbodies. The single blackbody fit discussed previously was held constant, while an extra component was added to fit the IRAS points shortward of the Herschel data. The FIR flux found in this way was around 25\% greater than that measured using only one blackbody. This figure is likely to be an overestimate of the true FIR flux, as the flux of the 12 \micron\ IRAS filter is dominated by PAH emission. For repeatability and simplicity, we adopted the single blackbody measurement with an uncertainty of 25\% on the derived total FIR fluxes. The total FIR fluxes for each object derived in this manner are listed in Table~\ref{tables:fluxes}.



\section{PDR Diagnostics}\label{sec:res}

It is clear from inspection of Table~\ref{tables:fluxes} that the [O~{\sc i}] 63 \micron\ line dominates cooling for both PDRs, dominating both of the other important cooling lines ([O~{\sc i}] 146 \micron\ and [C~{\sc ii}] 157 \micron) by factors of between five and eight. This is typical of dense PDRs. In addition CO rotational lines extend to very high J levels also indicating warm dense PDRs. On the other hand, the [C~{\sc ii}] and [C~{\sc i}] emission can also arise in low density gas illuminated by weak radiation fields. There is likely a mix of densities, temperatures, and filling factors that combine to produce the observed emission. We next use PDR line diagnostics to investigate the various phases.

\begin{table*}
\caption{\label{tables:diags}Results of Simple PDR Diagnostics$^a$}
\begin{center}
\renewcommand{\arraystretch}{1.5}
\begin{tabular}{p{6cm} l | p{1.1cm} p{1cm} | p{1.1cm} p{1cm}}
\hline\hline
Diagnostic & Figure &  log \Gnaught & log $n$ & log \Gnaught & log $n$ \\
           &        & \multicolumn{2}{c|}{IRAS~23133+6050} & \multicolumn{2}{c}{S~106} \\ 
\hline
\textit{General:} & & & & &\\
Spectral Types &  & 4.6 &  & < 4.9   &  \\
$\Sigma$ FIR   &  & 5.1 &  & 5.1--5.4 &  \\
 & & & & &\\

\textit{Low density} & & & & &\\
$\frac{\textrm{\cii\ 158 \micron}}{\textrm{\oi\ 63 \micron}}$ vs $\frac{\textrm{\cii\ 158 \micron\ + \oi\ 63 \micron}}{\Sigma \textrm{FIR}}$  & \ref{fig:diags2}    & 4.0 & 3.5 & 4.2 & 3.8 \\
$\frac{\textrm{\cii\ 158 \micron}}{\textrm{\oi\ 145 \micron}}$ vs $\frac{\textrm{\cii\ 158 \micron\ + \oi\ 145 \micron}}{\Sigma \textrm{FIR}}$& \ref{fig:diags2}    & 4.0 & 4.1 & 4.2 & 4.5 \\
$\frac{\textrm{\ci\ 370 \micron}}{\textrm{\ci\ 609 \micron}}$ vs $\frac{\textrm{\cii\ 158 \micron\ + \oi\ 63 \micron}}{\Sigma \textrm{FIR}}$  & \ref{fig:diags2}    & 3.2 & 2.5 & 3.6 & 3.2 \\
 & & & & &\\

\textit{High density} & & & & &\\
$\frac{^{12}\textrm{CO (J=7-6)}}{^{12}\textrm{CO (J=4-3)}}$ vs $\frac{\textrm{\cii\ 158 \micron\ + \oi\ 63 \micron}}{\Sigma \textrm{FIR}}$    & \ref{fig:diags3}    & 4.9 & 4.7 & 4.9 & 4.7 \\
$\frac{^{12}\textrm{CO (J=15-14)}}{^{12}\textrm{CO (J=7-6)}}$ vs $\frac{\textrm{\cii\ 158 \micron\ + \oi\ 63 \micron}}{\Sigma \textrm{FIR}}$  & \ref{fig:diags3}    & 5.2 & 5.6 & 5.2 & 5.6 \\
$\frac{^{12}\textrm{CO (J=15-14)}}{^{12}\textrm{CO (J=4-3)}}$ vs $\frac{\textrm{\cii\ 158 \micron\ + \oi\ 63 \micron}}{\Sigma \textrm{FIR}}$  & \ref{fig:diags3}    & 5.2 & 5.4 & 5.2 & 5.4 \\
 & & & & &\\

\textit{Questionable:} & & & & &\\
$\frac{\textrm{\ci\ 609 \micron}}{^{12}\textrm{CO (J=4-3)}}$ vs $\frac{\textrm{\cii\ 158 \micron\ + \oi\ 63 \micron}}{\Sigma \textrm{FIR}}$    & \ref{fig:diags4}  & 5.2 & 5.4 & 5.2 & 5.8 \\
\hline
\end{tabular}

\end{center}
\bigskip
$^a$: With \Gnaught\ expressed as a multiple of the Habing Field and $n$ in cm$^{-3}$.
\end{table*}


\subsection{FIR Continuum and Stellar Spectral Types}

The FIR continuum can be used to directly probe the UV radiation field impinging on the PDR system. The total energy absorbed by the dust in the system must be the same as that re-emitted. Previous authors (e.g., \citealt{1992ApJ...390..499M}) have used this relationship to calculate a value for \Gnaught. This calculation is summarized in Appendix B of \citet{1992ApJ...390..499M} where it is shown that the derived \Gnaught\ values weakly depend on the emitting structure. For a spherically symmetric source (e.g., IRAS23133+6050), it can be assumed that the absorbing and radiating surface areas are equal, and hence the relationship can be summarized as: \Gnaught\ = $\Sigma$FIR / 1 Habing field. For more complex sources the relationship between the radiating and absorbing surface areas is different, for example, \citealt{1992ApJ...390..499M} show that for a disk system that the derived \Gnaught\ will be different by a factor of two. S~106 would seem to be a mixture of these two paradigms, with a small disk around the central star being further encased within a molecular cloud/PDR, therefore there will exist a range of values of \Gnaught\ spanning these geometries. For each object the resulting values of \Gnaught\ assuming a spherically symmetric geometry are very similar at around 1.3$\times$10$^5$. For the disk region of S~106 though, \Gnaught\ could be up to 2.6$\times$10$^5$.

An additional measure of \Gnaught\ can be calculated using the distance from the ionizing source to the PDR and the stellar spectral type of the exciting star. \citet{2010pcim.book.....T} gives the relationship between \Gnaught\ and the stellar luminosity as: 

\begin{equation}\label{eq:gnaught}
G_0 = 625 \times \frac{L_{*}\chi}{4 \pi d^2}
\end{equation}

where $L_{*}$ is the stellar luminosity (in ergs s$^{-1}$), $\chi$ is the fraction of photon energy emerging from the \HII\ region between 6 and 13.6 eV, and $d$ is the distance between the star and the ionization front in cm. Both IRAS~23133+6050 and S~106 are powered by late O8-9 type stars, which have far UV luminosities around 3.25 $\times$ 10$^{4}$ times that of the sun\footnotemark. For IRAS~23133+6050, the ionized region has a radius of 4.5\arcsec, which at a distance of 5.5 kpc translates to 0.1 pc. This implies a \Gnaught\ value of around 5 $\times$ 10$^{4}$. Similarly for S~106, the radius of the ionization front is around 0.1 pc (based on a distance of 1.5 kpc and a radius of 12.5\arcsec; \citealt{2005AA...436..569N}), which implies a \Gnaught\ of about 8 $\times$ 10$^4$ when using the same assumptions. \citet{2005AA...436..569N} found that the \HII\ region of S~106 was approximately cylindrical, and as such the \Gnaught\ we calculate using their radius is only appropriate for the parts of the ionization front closest to the exciting star and therefore is actually an upper limit. In good agreement with our analysis, the Habing field for S~106 determined from the Herschel PACS 70 and 160 micron fluxes is 3 $\times$ 10$^4$ \Gnaught (value at a distance of 0.1 pc from S106 IR in a 12" beam, Schneider et al., in prep).

\footnotetext{ \citet{2003ApJ...584..797P} provide the relationship between stellar mass and the FUV flux, so the quoted figure, which represents $L_{*}\chi$ in Equation~\ref{eq:gnaught}, can be arrived at by assuming a stellar mass of around $\sim$21 M$_\odot$ for an O8-9 star.}


\subsection{Diagnostic Diagrams}

The measured line fluxes given in Table~\ref{tables:fluxes} can be combined in various ways to produce ratios which are sensitive to the physical conditions in the PDR. The variation of pairs of these ratios with respect to the fundamental PDR parameters \Gnaught\ and $n$\footnotemark are shown in Figs. \ref{fig:diags2} and \ref{fig:diags3}. To construct these diagrams we used the results of the PDR models of \citet{2010ApJ...716.1191W} and \citet{2012ApJ...754..105H}. There are a wide variety of different PDR models available in the literature, and they have been extensively compared and validated against each other, for example see \citet{2007A&A...467..187R}, in which the model we have chosen to employ here is referred to as the HTBKW model. These models have all been shown to reasonably agree for most PDR diagnostics and all reproduce the same trends when run on idential data \citep{2007A&A...467..187R}.

In each case we consider a range of \Gnaught\ values between 10$^3$ and 10$^6$ Habings and a range of densities between 10$^3$ and 10$^7$ cm$^{-3}$. In Table~\ref{tables:diags} we give an overview of the physical conditions indicated by the various diagnostics discussed in addition to those implied by the total FIR flux and the stellar spectral types of both sources.

\footnotetext{Henceforth we use $n$ to refer to the density of hydrogen nuclei such that for a purely molecular gas n(H$_2$) = 0.5 $\times$ n.}


\subsubsection{[C~\sc{ii}] and [O~\sc{i}]}

\begin{figure}
	\begin{center}
	\includegraphics[width=7.25cm]{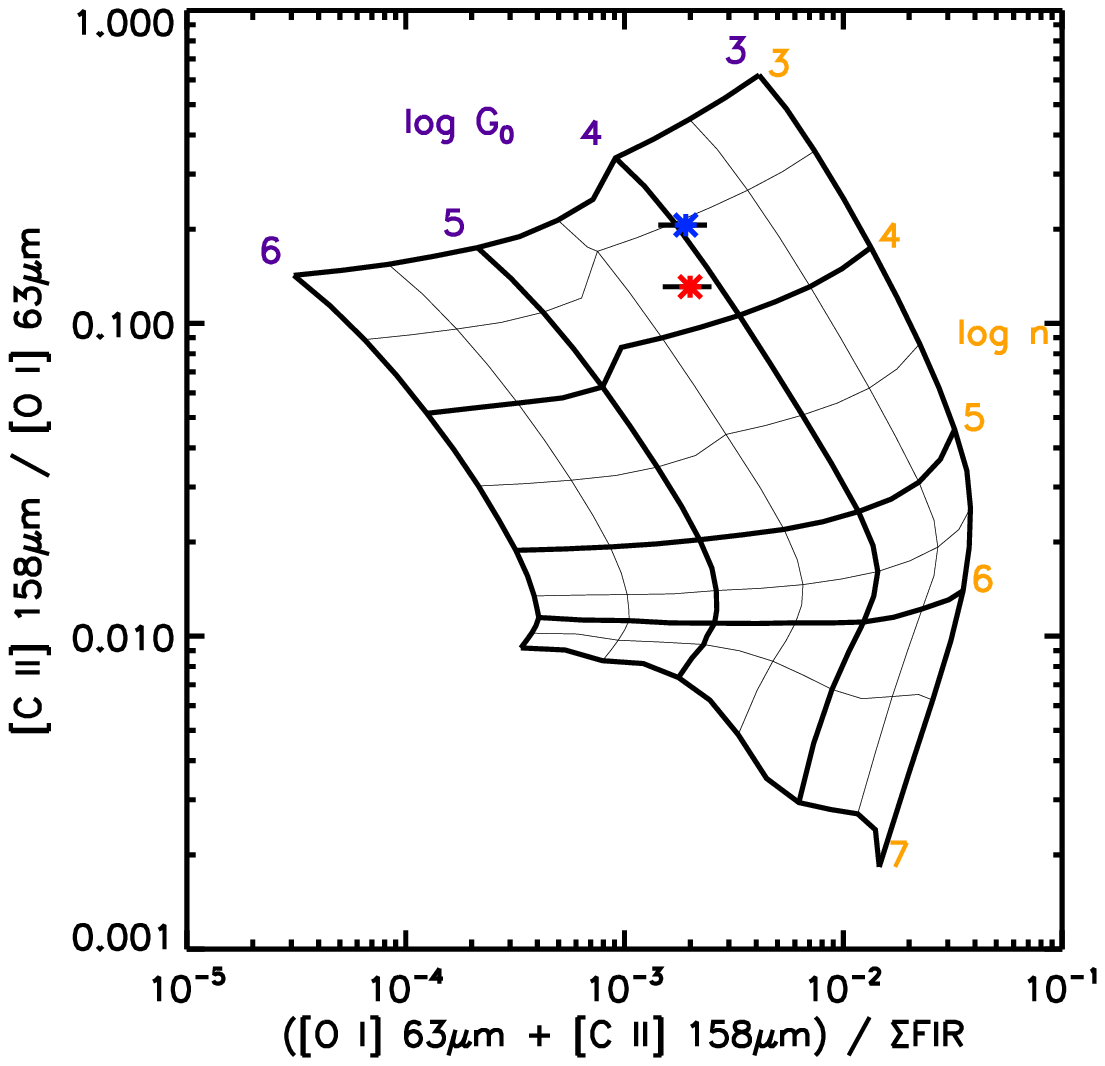}
	\includegraphics[width=7.25cm]{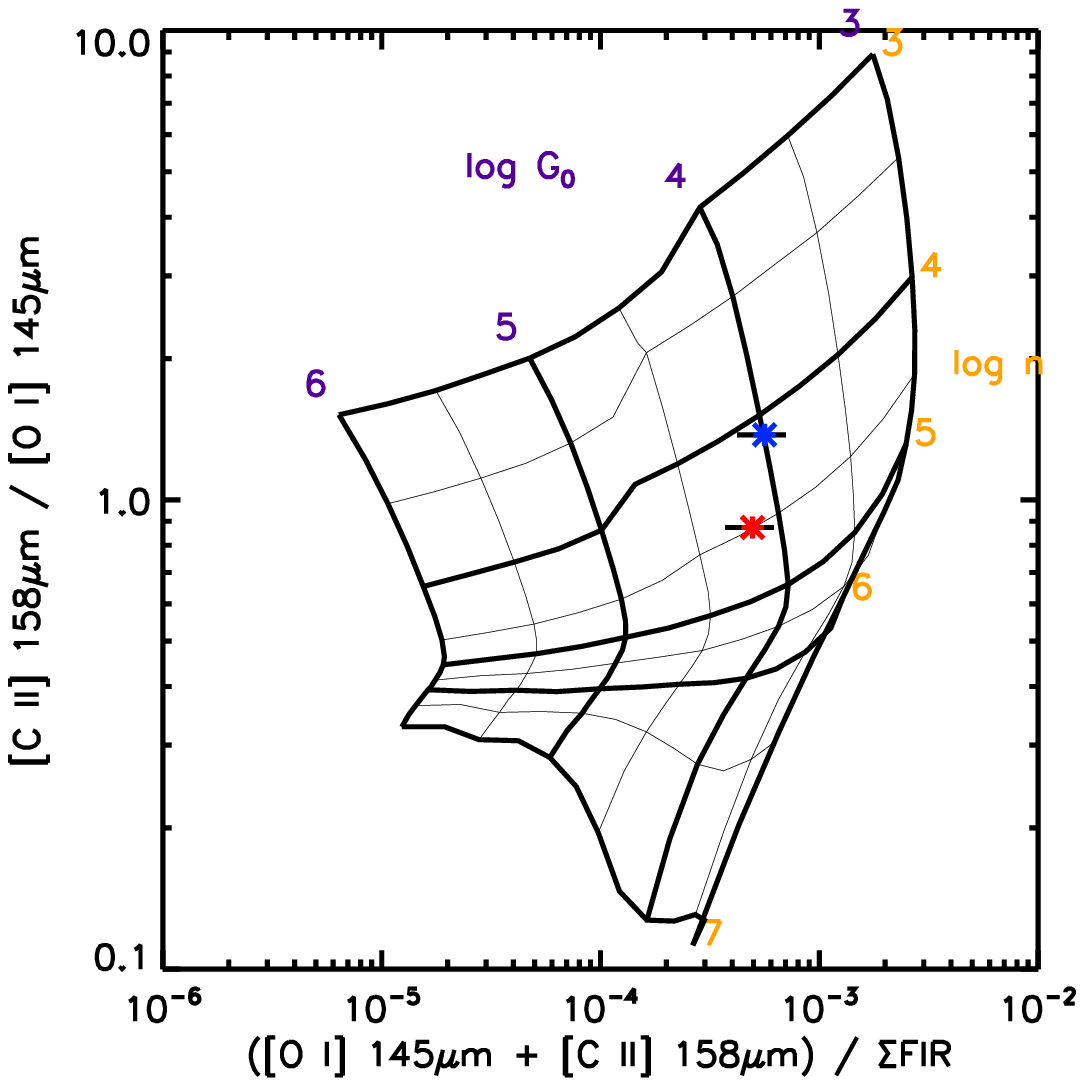}
	\includegraphics[width=7.25cm]{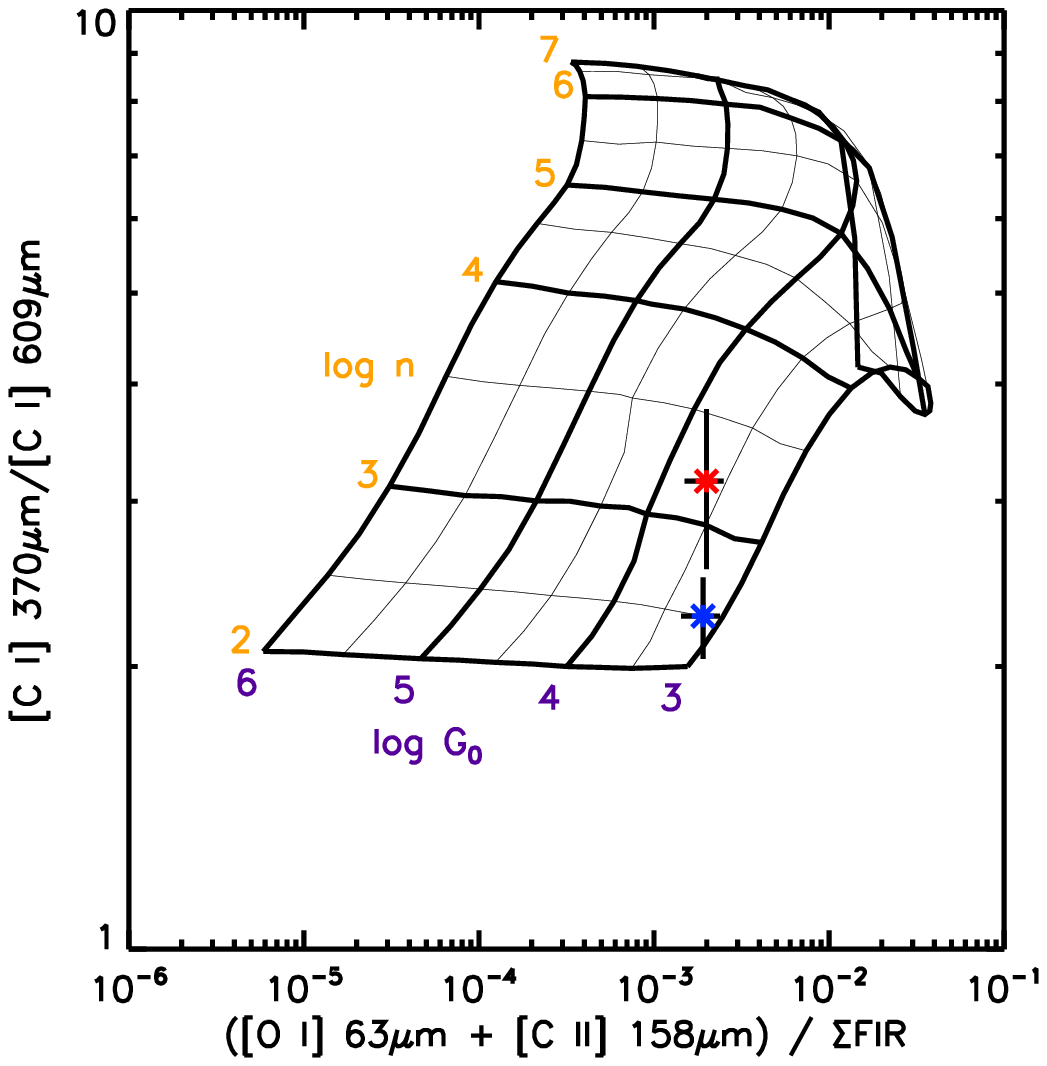}
	\end{center}
	\caption{ PDR diagnostic diagrams for observed fine-structure lines -- based on the PDR models of \citet{2010ApJ...716.1191W} and \citet{2012ApJ...754..105H}. \textit{Top:}  \cii\ 158 \micron\ / \oi\ 63 \micron\ versus (\oi\ 63 \micron\ + \cii\ 158 \micron)/ $\Sigma$FIR. \textit{Middle:} \cii\ 158 / \oi\ 145 versus (\oi\ 145 \micron\ + \cii\ 158 \micron)/ $\Sigma$FIR.  \textit{Bottom:} \ci\ 370 / \ci\ 609 versus (\oi\ 145 \micron\ + \cii\ 158 \micron)/ $\Sigma$FIR with the density grid extended down to log n = 2. In each figure the blue and red points represent the values found for IRAS~23133+6050 and S~106 respectively.}
	\label{fig:diags2}
\end{figure}

The primary diagnostics usually employed for PDRs involve ratios between FIR fine-structure cooling lines. Usually the ratios involving \cii\ 158 \micron\ and the two \oi\ lines are employed reflecting that these lines are the best understood of those included in PDR models. Typically such ratios do not correspond to unique PDR parameters, as their dependence of the line ratios on \Gnaught\ and $n$ is complex. In order to produce a single value, other diagnostics must be employed such that intersections at specific \Gnaught\ and $n$ values are found. The simplest combination of diagnostics which yields such physical conditions occurs when considering \cii\ 158 \micron\ / \oi\ 63 \micron\ and (\oi\ 63 \micron\ + \cii\ 158 \micron) / $\Sigma$FIR diagnostic diagram (Fig.~\ref{fig:diags2}, top). 

This diagram implies that both PDRs have similar physical conditions, with log \Gnaught\ around 4 and log $n$ around 3.75, well within typical PDR conditions. These values are compatible with those which were discussed in Section~\ref{sec:targs}. For IRAS~23133+6050 the density of $\sim10^4$ agrees nicely with that expected given the density of the \HII\ region, while for S~106 this density likely represents the lower density material which makes up the majority of the PDR volume (e.g., \citealt{2003AA...406..915S}). 

The ratio of the \oi\ lines measured for both sources is of the order of \oi\ 145 \micron\ / \oi\ 63 \micron\ $\simeq$ 0.15, which agrees with the measurements of \citet{2003AA...406..915S} who found 0.17 using ISO-LWS observations. This clearly indicates that one or both of the \oi\ lines is optically thick (e.g., \citealt{1985ApJ...291..722T}; Fig. 2): if both lines were optically thin their ratio would be less than 0.1. More recently it has been shown that in many cases the \oi\ line ratio indicates optically thick conditions, though other explanations for such ratios exist. It was concluded by \citet{2006A&A...446..561L} that self absorption of the \oi\ 63 \micron\ line along the line of sight with modest optical depths of around 1--2 could be responsible for high values of the \oi\ 145 \micron\ /  \oi\ 63 \micron\ ratio\footnotemark. Those results refer mainly to an ISO-LWS sample of dark clouds, for which the density required for both \oi\ lines to be optically thick was considered very unlikely. For IRAS~23133+6050 and S~106 though, such densities may be plausible -- however \oi\ 63 \micron\ self absorption would likely also play a role in shaping the flux ratio, rendering it difficult to disentangle the two.

\footnotetext{\citet{2006A&A...446..561L} actually consider the inverse of this quantity and as such their higher/lower than adjectives are switched, we adopt the same sense as is provided by the PDRT diagnostics of \citet{2006ApJ...644..283K, 2008ASPC..394..654P}.  }


\subsubsection{[C~\sc{i}]}

In addition, we also investigate the \ci\ fine-structure lines by considering the ratio of the fluxes of the 370 \micron\ line over the 609 \micron\ line (bottom panel of Fig.~\ref{fig:diags2}). This ratio indicates very different physical properties to those mentioned previously, with the ratio of the \ci\ lines indicating low density, weak radiation field PDR conditions. It should be noted though that it has been shown that the \ci\ intensities given by models are generally uncertain (maximal variation of around a factor of ten; \citealt{2007A&A...467..187R}) and therefore the results for this line ratio are highly model dependent.  


\subsubsection{$^{\textrm{12}}$CO}\label{sec:CO}

\begin{figure}
	\begin{center}
	\includegraphics[width=7.25cm]{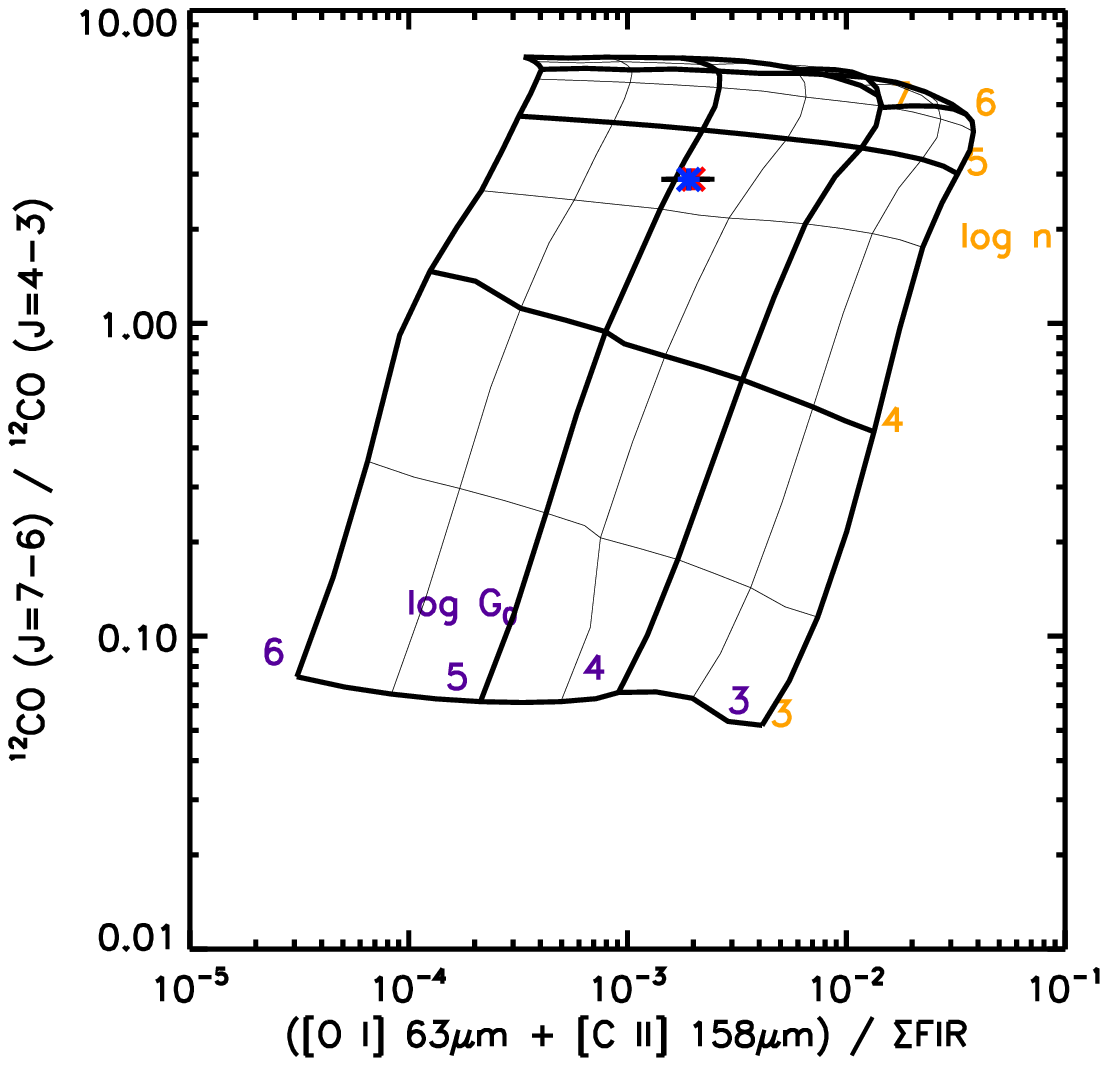}
	\includegraphics[width=7.25cm]{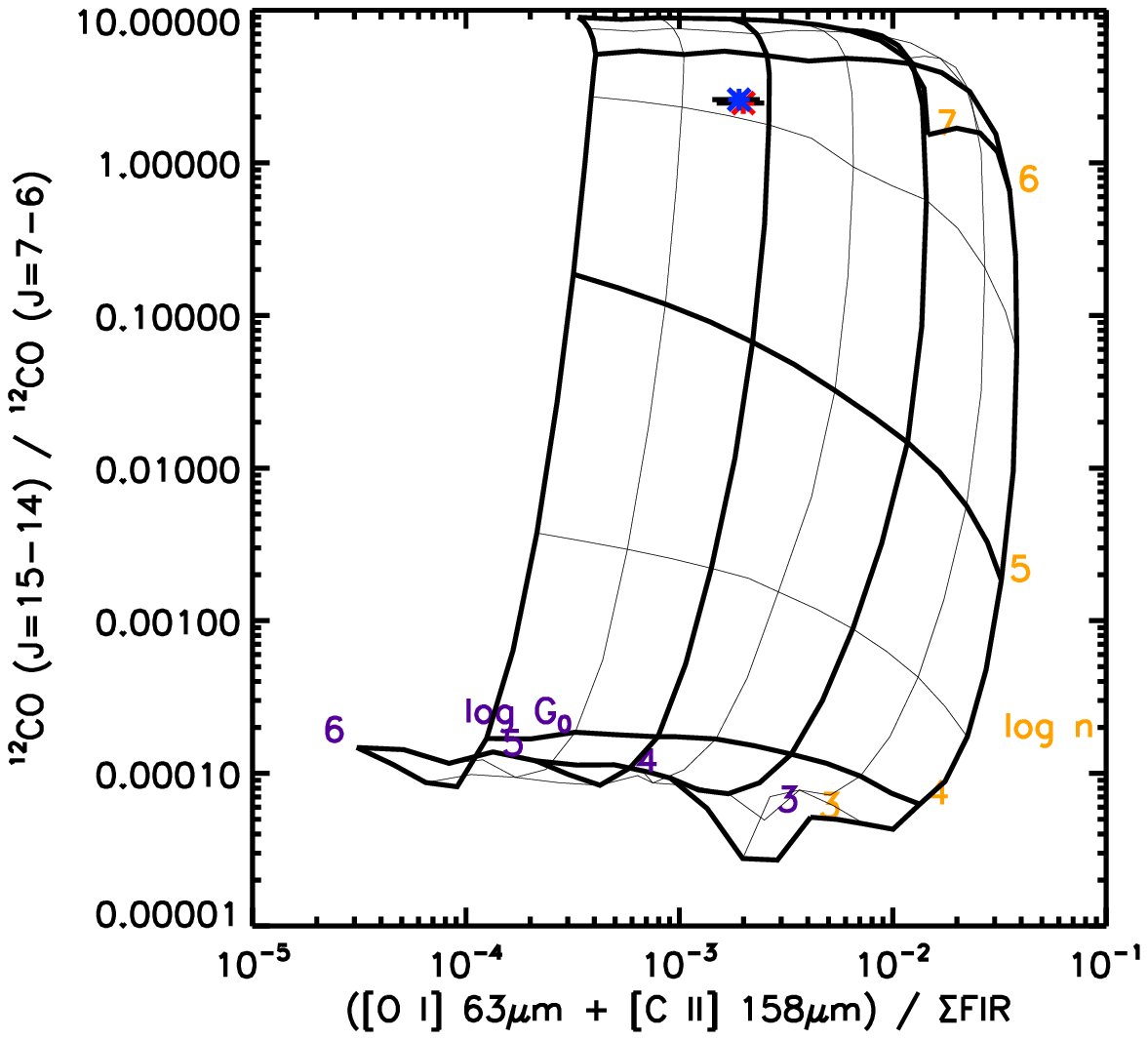}
	\includegraphics[width=7.25cm]{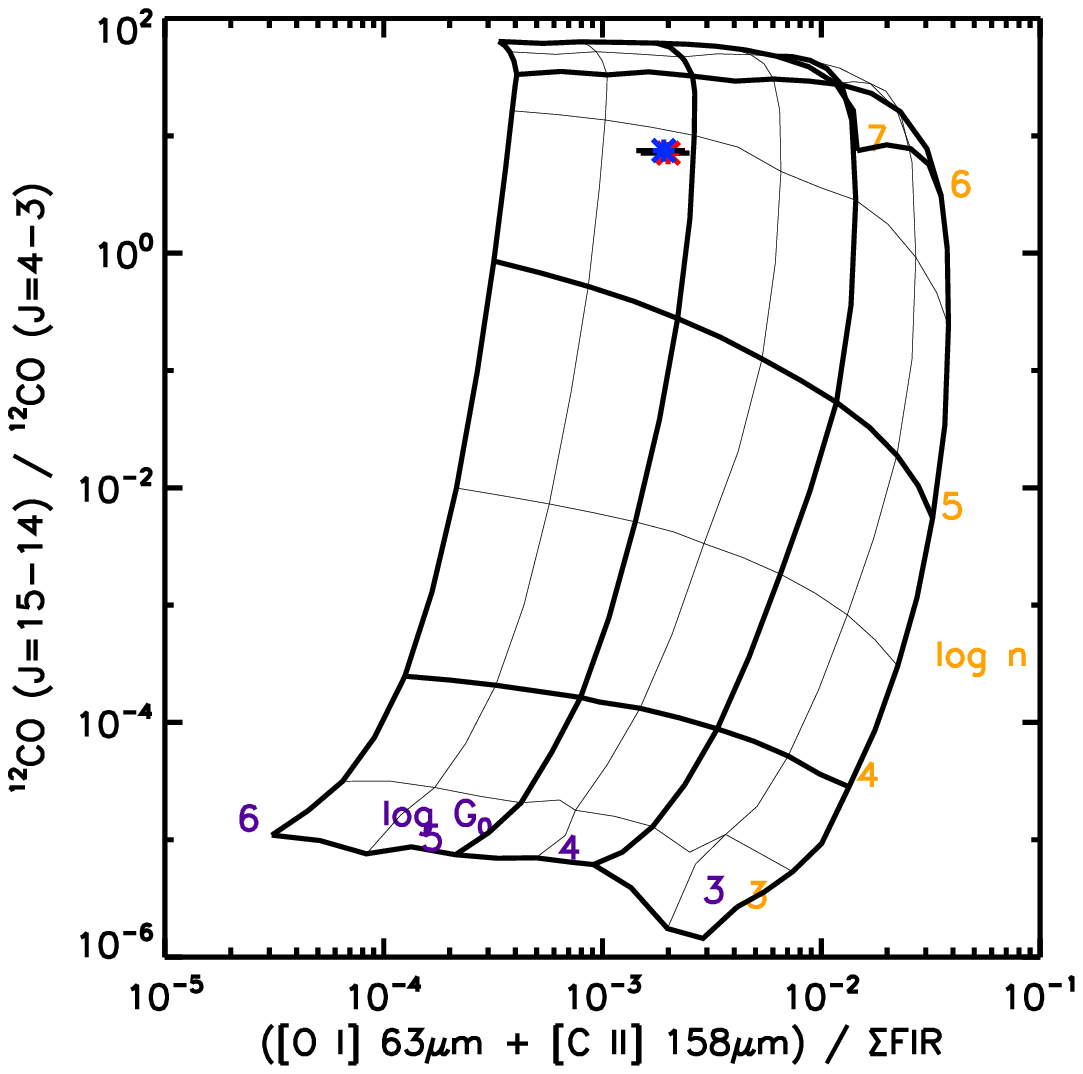}
	\end{center}
	\caption{ PDR diagnostic diagrams for ratios of $^{12}$CO(J=15-14), $^{12}$CO(J=7-6) and $^{12}$CO(J=4-3) versus (\oi\ 63 \micron\ + \cii\ 158 \micron)/ $\Sigma$FIR.  In each figure the blue and red points represent the values found for IRAS~23133+6050 and S~106 respectively. The data points indicate a much denser environment with a stronger UV field than that seen for the fine-structure line diagnostics.}
	\label{fig:diags3}
\end{figure}

\begin{figure}
	\begin{center}
	\includegraphics[width=7.25cm]{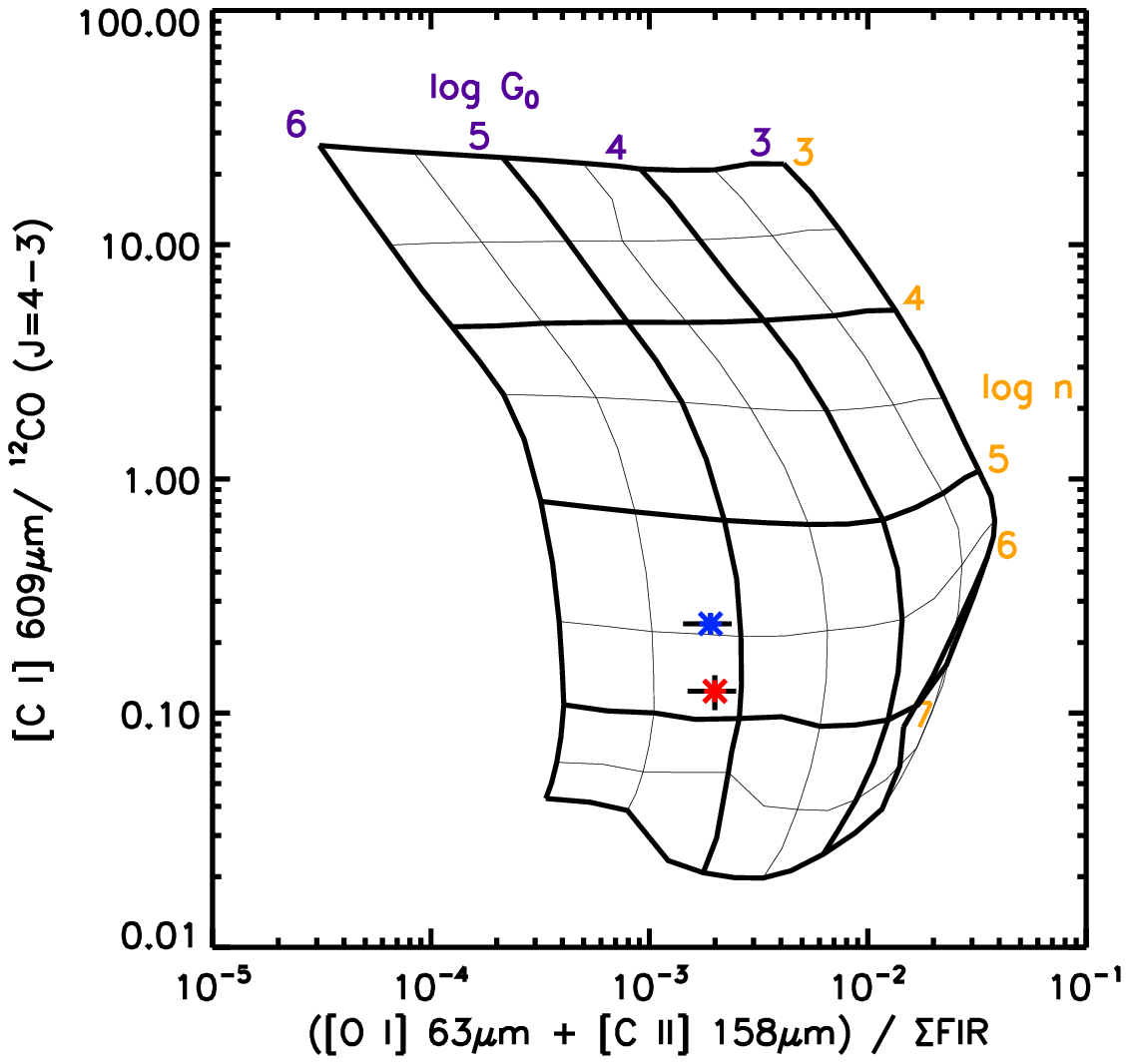}
	\end{center}
	\caption{ \ci\ 609 \micron\ / $^{12}$CO (J=4-3) versus (\oi\ 63 \micron\ + \cii\ 158 \micron)/ $\Sigma$FIR PDR diagnostic diagram. The blue and red points represent the values found for IRAS~23133+6050 and S~106 respectively.}
	\label{fig:diags4}
\end{figure}

In addition to the fine-structure cooling line diagnostics we can also employ the properties of the molecular lines in the spectra in the same way. \citet{1999ApJ...527..795K} provide the various diagrams for the different $^{12}$CO transitions, here we investigate ratios of three: 4-3, 7-6 and 15-14, but using the updated PDR models of \citet{2010ApJ...716.1191W} and \citet{2012ApJ...754..105H}.

The ratio of the two mid-J lines: $^{12}$CO(J=7-6)/$^{12}$CO(J=4-3) is shown in Fig.~\ref{fig:diags3} (top), and indicates a much stronger UV field (log \Gnaught\ = 4.9) and denser environment (log $n$ = 4.8) for both sources than the conditions indicated by the fine-structure lines. For the highest J $^{12}$CO line, J=15-14, even denser and more strongly illuminated conditions are found, with densities an order of magnitude higher and radiation fields around double that found from the lower excitation lines. Diagnostic diagrams for the $^{12}$CO (J=15-14) line, as compared to the lower excitation CO lines, are shown in the lower two panels of Fig.~\ref{fig:diags3}.

In addition to the pure $^{12}$CO ratios, we also show the ratio of the \ci\ fine-structure line with the $^{12}$CO(J=4-3) line in Fig.~\ref{fig:diags4}. This diagnostic clearly shows a similar pattern to that of the other $^{12}$CO diagnostic ratios, i.e. high log \Gnaught\ ($\sim$ 5) and log $n$ ($\sim$ 5).


\subsection{CO Ladder fluxes}

\begin{figure}
	\begin{center}
	\includegraphics[width=9cm]{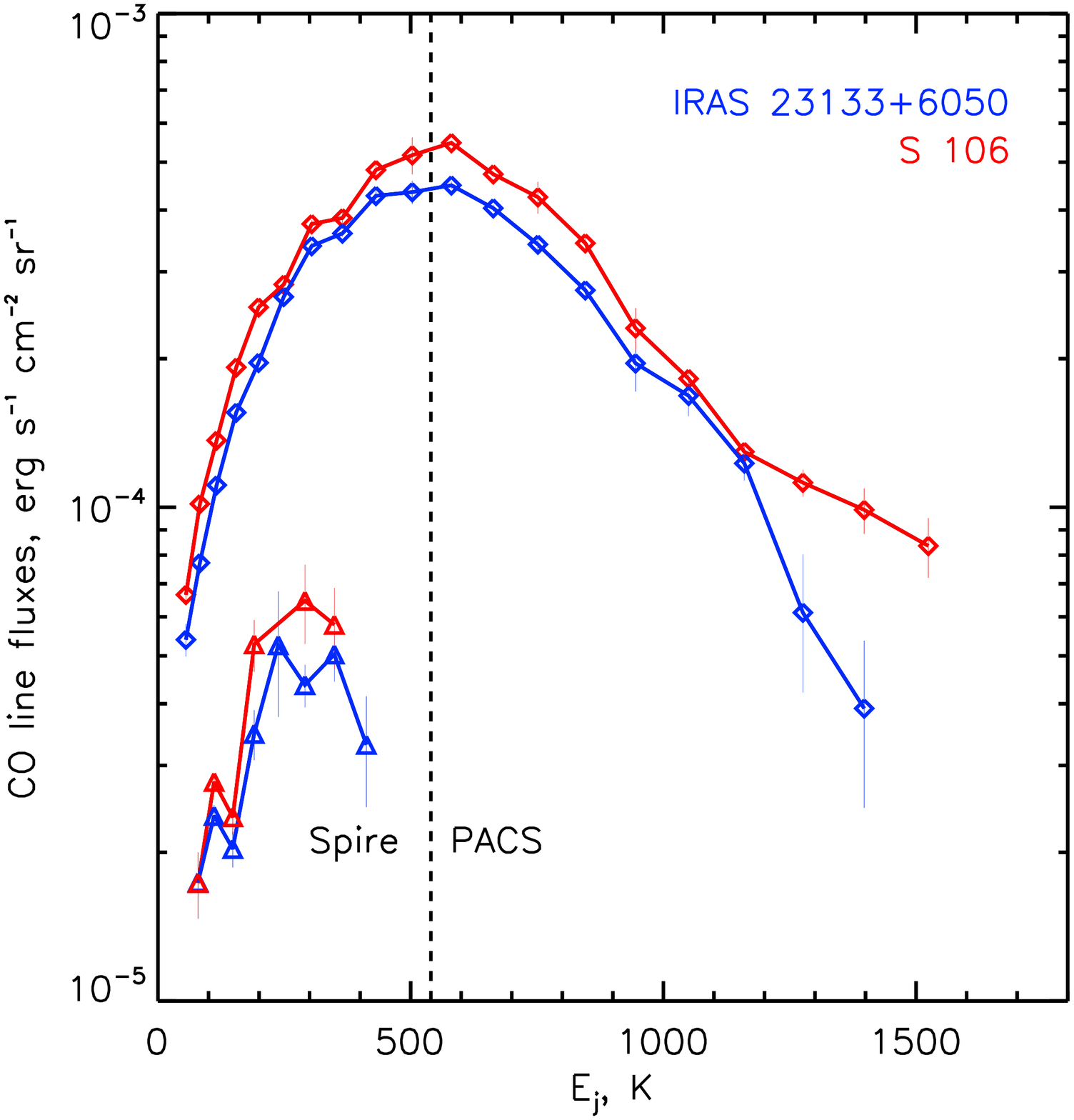}
	\end{center}
	\caption{$^{12}$CO (squares) and $^{13}$CO (triangles) ladder fluxes for IRAS~23133+6050 and S~106 in terms of the excitation energy of that level.   }
	\label{fig:ladflux}
\end{figure}

In the previous section we exploited the simple relationships between different CO transitions resulting from PDR models to find the physical conditions. It is also possible to derive such conditions directly from the fluxes of the whole set of molecular lines. In Fig.~\ref{fig:ladflux} we show the $^{12}$CO and $^{13}$CO line fluxes against the energy of the upper level E$_J$. The detections of the $^{12}$CO lines for both IRAS~23133+6050 and S~106 are measured across the entire SPIRE/PACS range, that is, from J$_u$ = 4 to 23, while for $^{13}$CO the range is smaller, from J$_u$ = 5 to 12, owing to their weaker fluxes. The $^{12}$CO fluxes shown in Fig.~\ref{fig:ladflux} (a) are remarkably similar for both objects, peaking at around E$_j$ = 550 K, or J$_u$ = 13 or 14. The only difference between IRAS~23133+6050 and S~106 comes from the higher J lines, where for S~106 the four highest J lines show additional excitation, whilst those for IRAS~23133+6050 continue to decrease. The possible implications of this will be discussed in the following section. The $^{13}$CO line fluxes seem to peak at around E$_j$ = $\sim$300 K, or J$_u$ = 10, however the data are noisy and it is not clear that this apparent peak is real.

\begin{figure}
	\begin{center}
	\includegraphics[width=9cm]{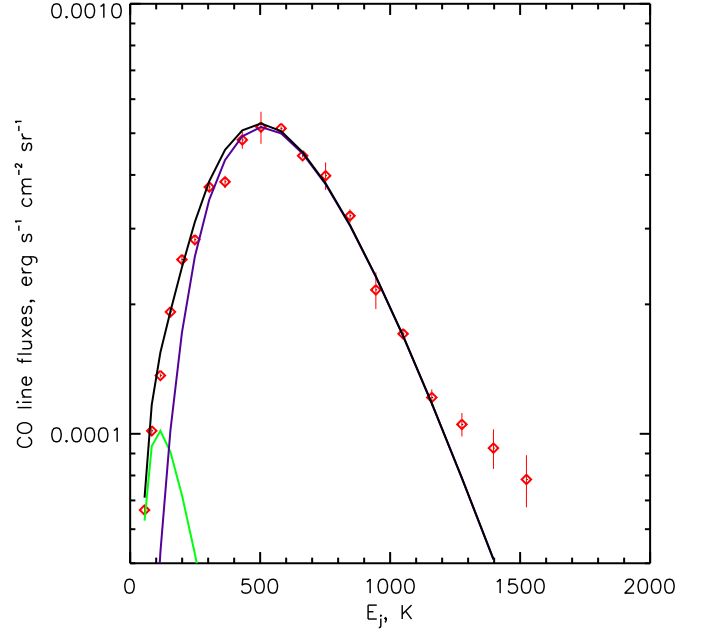}
	\end{center}
	\caption{Combination of RADEX models scaled to fit the S~106 $^{12}$CO ladder. The low density component with log $n$ = 4, T = 300 K is represented by the green line, while the dominant contribution arises from the dense component with log $n$ = 5.6 and a temperature of 400 K which is represented by the purple line. The aggregate model (black line) matches the observed $^{12}$CO ladder from J$_U$ = 4 to 20.}
	\label{fig:radex}
\end{figure}

The peak of the $^{12}$CO fluxes is itself an indicator of the physical conditions of both sources. Using the density of the high density component predicted by the simple diagnostics (n $\sim$ 10$^{5.6}$~cm$^{-3}$) and the radiative transfer code \verb+RADEX+ \citep{2007A&A...468..627V} we calculated the fluxes of $^{12}$CO lines for a variety of temperatures. The required temperature to fit the peak is then around 400 K. In Fig~\ref{fig:radex} we show this model, combined with a model representing the low density component at the same temperature, which have been scaled to match the overall flux of the whole CO ladder. However, the combination of density and temperature which fits the peak is not unique, and higher or lower density models can fit the peak almost equally well with suitably reduced or increased temperatures. Our choice of n = 10$^{5.6}$~cm$^{-3}$ is driven by the simple diagnostics discussed in previous subsections. In contrast, the highest and lowest `plausible' solutions considered have densities and temperatures of 10$^7$ or 5$\times$10$^4$ cm$^{-3}$ and 200 or 1500 K respectively. Thus \verb+RADEX+ models lead us to conclude that the most plausible explanation for the mid-J (J$_U$ $\simeq$ 6--20) $^{12}$CO ladder is a (very) high density PDR with a temperature of around 200--400 K. We attempted to fit the high-J S~106 lines using \verb+RADEX+, however it was not possible to find a good fit to those points without severely compromising the quality of the fit to the other points. 


\subsubsection{CO Rotation diagrams}\label{sec:radex_rot}

\begin{figure}
	\begin{center}
	\includegraphics[width=9cm]{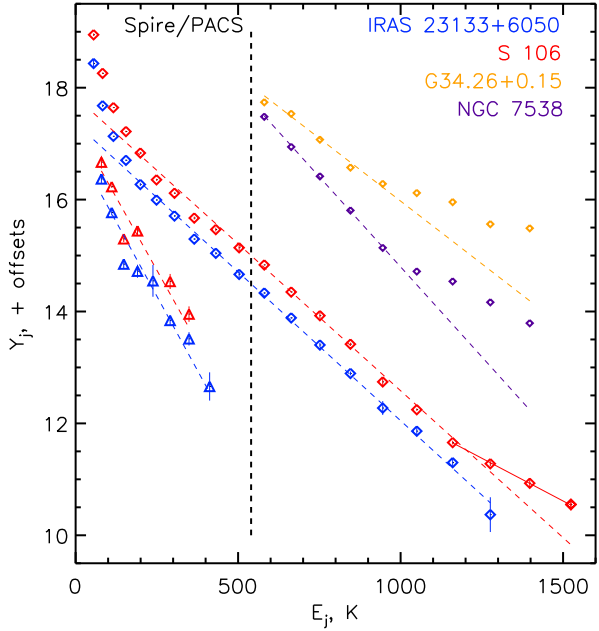}
	\end{center}
	\caption{$^{12}$CO (diamonds) and $^{13}$CO (triangles) rotation diagrams for IRAS~23133+6050 and S~106. Two ultra-compact \HII\ regions from the \citet{2014A&A...562A..45K} sample which display the warm CO component are also shown as orange, purple lines. Simple fits to the lower J CO lines are shown to demonstrate the deviation of the higher J CO levels from this trend for S~106 and the Karska sources. A fit to the highest J CO lines for S~106 is also included. The Karska sources have been offset in $y$ for clarity. The first $^{12}$CO and $^{13}$CO points for both of our sources are 4-3 and 5-4 respectively.}
	\label{fig:rotdiags}
\end{figure}

To further illustrate the differences between our sources and those found in the literature we create rotation diagrams from the $^{12}$CO and $^{13}$CO fluxes (Fig.~\ref{fig:rotdiags}). Following \citet{2000A&A...360.1117J} we construct rotation diagrams showing the relationship between the line fluxes $F_{ji}$, the temperature $T_x$ and the column density $N_0$ as follows:

\begin{equation}
E_j / kT_x = ln\left( \frac{N_0}{\Phi(T_x)d^2}\right) - ln\left(\frac{4\pi F_{ji}}{A_{ji}h\nu_{ji}g_j} \right),
\end{equation}

where $g_j$ refers to the statistical weight of the upper level $j$, $\nu_{ji}$ is the frequency of the transition, $\Phi$($T_x$) is the partition function, A$_{ji}$ is the oscillator strength for the particular transition and d is the distance to the source. \citet{2000A&A...360.1117J} refer to the second term as $Y_j$ and we follow this convention. The temperature and column densities can then be found in the limit of optically thin emission in local thermodynamic equilibrium (LTE) by fitting a straight line to the $E_j$, $Y_j$ coordinates. There are many potential problems with this method, mainly involving the aforementioned assumptions (see, for example, \citealt{1999ApJ...517..209G}). In light of that, it is not surprising that both IRAS~23133+6050 and S~106 do not follow exactly straight lines. Nevertheless, since it is a common procedure, we will analyze the observations using this technique and compare to other methods and models. 

The curvature exhibited at low E$_j$ is expected and usually interpreted as requiring a second, warm, temperature component (c.f. Figure~\ref{fig:radex}). These levels trace the warmish material on the surface of the (putative) molecular cloud, the colder molecular cloud material would only manifest itself in unobserved lower J transitions (J < 3). However recent work has shown that this curvature can be explained in the optically thin limit by deviations from LTE at densities below 10$^{5}$ cm$^{-3}$ and constant temperatures \citep{2012ApJ...749..125N}. This effect arises because, in the low density limit, the population is dominated by the low lying transitions, with collisional excitation and spontaneous decay balanced for the higher level transitions. 

From these best fits to the $^{12}$CO lines of S~106 and IRAS~23133+6050 we find excitation temperatures of around 190 K and around 10$^{49}$ molecules along the line of sight (summarized in Table~\ref{tables:rot_dia_fits}). We have also included a linear fit to the high excitation component of the S~106 $^{12}$CO spectrum -- this fit yields a much higher excitation temperature of around 330 K with a lower density. The derived excitation temperatures for $^{12}$CO are systematically lower than those found by the \verb+RADEX+ models discussed in the previous subsection. This discrepancy is likely due to the assumptions about optical depth which underpin the use of rotation diagrams -- it seems unlikely that many of the low J CO lines are optically thin, for example\footnotemark. The \verb+RADEX+ models make no such assumptions and are therefore a somewhat more trustworthy indicator of the temperature of the various components, on the other hand the \verb+RADEX+ models assume either an arbitrary gas temperature distribution or constant temperature, and constant molecular abundances. However, we could not find a plausible solution for the very high excitation S~106 points with \verb+RADEX+ models, whereas the rotation diagrams tell us that the temperature of that component is higher than that of the medium density component by at least 100 K. However, even this argument relies on the assumption that our sources are in LTE, which is not altogether clear from inspection of Figure~\ref{fig:rotdiags}. As such the parameters of the rotation diagram fits should be regarded with some caution, we will discuss how well these results compare to the results of the more detailed models in Section~\ref{sec:discCO}.

\footnotetext{A simple estimate of the $^{12}$CO optical depth can be made from the ratio, R, of the $^{12}$CO / $^{13}$CO line intensities. The optical depth, $\tau_{12}$ = X/R where X is the isotope abundance ratio \citep{2013PASA...30...44B}. For plausible values of X (e.g. see \citealt{2005ApJ...634.1126M}), the $^{12}$CO line is optically thick to the highest J level for which both $^{12}$CO and $^{13}$CO lines are available. }

\begin{table}
\caption{\label{tables:rot_dia_fits} Properties of fits to Rotation Diagram Components}
\begin{center}
\begin{tabular}{l c c c c}
\hline\hline
Source & Species & J$_\mathrm{U}$'s & T [K] & log(N)  \\
\hline
\\
IRAS 23133+6050 & $^{12}$CO & all & 188 & 49.33 \\
IRAS 23133+6050 & $^{13}$CO & all & 94  & 48.84 \\
\\
S~106 		& $^{12}$CO & 4-20  & 190 & 48.91 \\
S~106 		& $^{12}$CO & 20-23 & 327 & 48.00 \\
S~106 		& $^{13}$CO & all & 96  & 48.40 \\
\hline
\end{tabular}

\end{center}
\end{table}

The linear component seen in the four highest J $^{12}$CO lines of S~106 betrays the possible existence of an additional physical component. There appear to be two possible explanations for this phenomenon: an even higher density PDR; or shocks in the S~106 environment. Such additional excitation in the upper CO levels has been observed commonly in a related class of objects: protostars. We have therefore included in the rotation diagram several prominent examples from a recent survey of high mass star forming regions using PACS by \citet{2014A&A...562A..45K}. The two objects (G34.26+0.15 and NGC 7538) were chosen for comparison on the grounds that they are ultra-compact \HII\ regions and therefore closest in structure to S~106 and IRAS~23133+6050. In fact, NGC 7538 is actually a bipolar \HII\ region similar to S~106. It is clear that these sources display an extra component, similar to that seen for S~106. In order to visualize the offset caused at high-J, we have included best fit lines to the mid-J linear portion of the data for each object in Fig.~\ref{fig:rotdiags}. In high mass protostars such as those studied by \citet{2014A&A...562A..45K} the additional excitation is thought to arise from a combination of shocks and PDR emission (for low mass protostars such excitation is usually thought to be dominated by shocks, e.g., \citealt{2010A&A...518L.121V}). These possibilities will be discussed in Section~\ref{sec:discCO}. In agreement with \citet{2000AA...358.1035V}, we conclude that the majority of the CO emission observed towards S~106 is due to a dense PDR, particularly for the low to mid J lines, however it is clear that the high J CO lines may be indicative of a contribution by a shock. 


\subsection{Summary of Diagnostics}

The simple diagnostics reveal two very similar sources, with subtle differences possibly driven by the different physical structures of the two sources. Both show high \Gnaught, as revealed by the total IR fluxes, but with lower radiation fields indicated by the various fine-structure and molecular line ratios. Thus a two component model is required, where the PDR is comprised of two distinct regions: i) a hot (T $\sim$ 400 K), dense region characterized by high \Gnaught\ ($>$10$^5$) and high density ($>$10$^5$ cm$^{-3}$) traced by the various $^{12}$CO lines; and ii) a less dense region region dominating the emission of the atomic cooling lines characterized by moderate UV fields (\Gnaught\ $\sim$10$^4$) and densities (10$^4$ cm$^{-3}$) with a temperature of around 300 K. The presence of such a distinction is not a surprise however, as models of PDRs have included density contrasts of this order since the early nineties (e.g., \citealt{1990ApJ...365..620B}) and it has been observed in a number of objects, from Orion to M17 (e.g., \citealt{1990ApJ...356..513S, 1994ApJ...422..136T}). In addition the very high excitation CO lines for S~106 show an even hotter component, which may arise from a shock or a much hotter PDR than those considered thus far. A detailed comparison between our sources and typical galactic PDRs will be presented in Section~\ref{sec:comp}.


\section{Numerical PDR models}\label{sec:num}

The densities and UV fields found in the previous section can be regarded as line of sight averages as they do not take into account the geometry of the system or the absolute fluxes of the lines. In order to extract extra information from the observations, we used the simple analysis as the input for numerical PDR codes. This process should help illuminate the structure along the line of sight producing the emission observed.

We use the PDR model described by \citet{2010ApJ...716.1191W}, but modified to use either constant density or constant thermal pressure. The model builds on that of \citet{2006ApJ...644..283K}, which in turn combined the initial PDR models of \citet{1985ApJ...291..722T} and \citet{1990ApJ...358..116W}. In addition the model includes both the molecular hydrogen processes incorporated into the Meudon PDR code (e.g., \citealt{2006ApJS..164..506L}) and molecular freeze-out and oxygen chemistry as described by \citet{2012ApJ...754..105H}. The model assumes a plane parallel slab geometry illuminated from one direction with a specified 1D UV field and calculates the thermal balance and abundances of the dominant atomic and molecular species as well as their line emission. These models have been used to interpret a variety of observations of molecular and atomic species in a wide range of physical conditions including OH$^+$, H$_2$O$^+$, and H$_3$O$^+$ in diffuse gas \citep{2012ApJ...754..105H}, H$_2$O and HF in diffuse gas (\citealt{Sonnentrucker}), [C~{\sc i}] and CO in low UV field molecular clouds \citep{2014ApJ...782...72B, 2014ApJ...784...80L}, H$_2$ in dense PDRs \citep{2011ApJ...741...45S} and [C~{\sc ii}] and [O~{\sc i}] in dense PDRs and extragalactic nuclei \citep{2006ApJ...644..283K, 2012ApJ...747...81C}. 

\begin{figure*}
	\begin{center}
	\includegraphics[width=18cm]{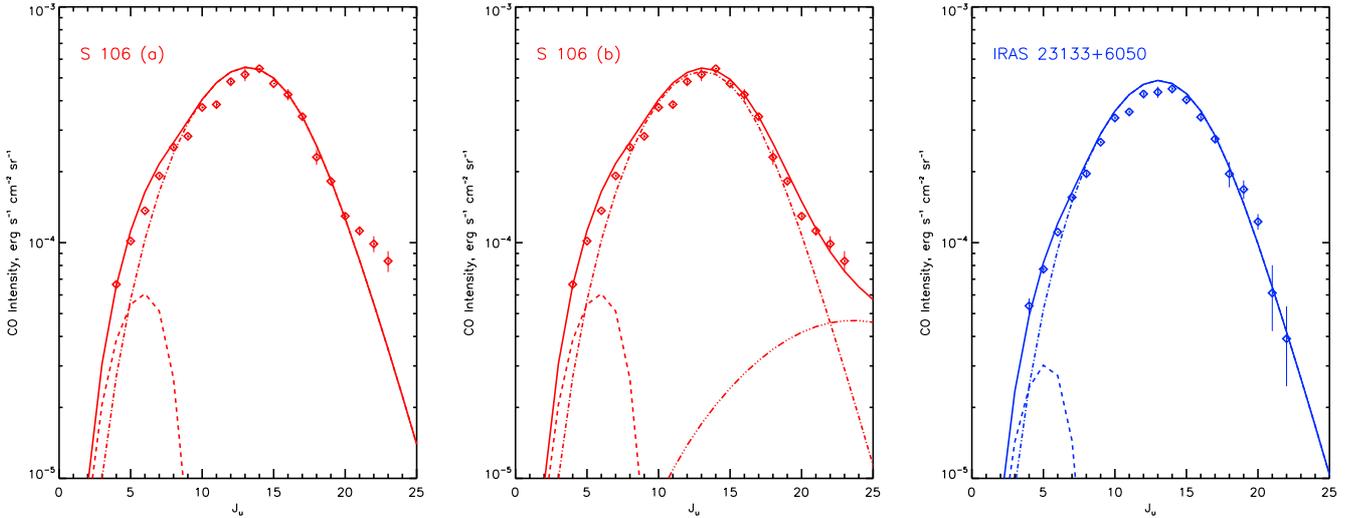}
	\end{center}
	\caption{PDR model fits to $^{12}$CO ladders. (a) Two component PDR model for S~106; (b) three component PDR model for S~106; (c) two component PDR fit for IRAS~23133+6050. In each panel the dashed line represents the low pressure component and the dot-dashed lines the high pressure component. In the centre panel the final, very high density PDR component for S~106 is represented by a triple-dot-dashed line.} 
	\label{fig:modelfit}
\end{figure*}

In order to get good fits to the observations of both PDRs constant thermal pressure models were used, with one exception which will be discussed shortly. In each case the final model shown in Fig.~\ref{fig:modelfit} is a combination of high pressure and low pressure models with the parameters guided by the properties of each regime from the simple diagnostics. Because the simple diagnostics report only average properties within the beam, it is necessary to scale the high and low density models for each source to match the overall fluxes. Hence we scaled the models using the CO line fluxes and found good agreement with the exception of the highest J lines for S~106 which display additional excitation as discussed earlier. In order to attempt to fit these lines, a third, extremely high density, constant density PDR model component was introduced\footnotemark. The degree to which each component needs to be scaled can be interpreted as the filling factor for the high density model, and a rough estimate of the number of PDRs on the line of sight for the low density model. 

\footnotetext{We refer to the two component model as S~106 (a) and the three component model as S~106 (b).}

Since our models are for constant thermal pressure, both the temperature and density vary with depth within the cloud. We characterize the physical conditions where the line emission arises at the depth where one half of the intensity is produced. For IRAS 23133+6050, the peak CO line emission (CO 13-12) comes from the high pressure component at T = 217 K, n = 1.3$\times$10$^6$ cm$^{-3}$. The [O~{\sc i}] 63 \micron\ line comes from a region closer to the cloud surface where temperatures are higher and densities are lower. We find T = 2030 K, and n = 7.6$\times$10$^4$ cm$^{-3}$. The low pressure component provides about half of the observed CO 5-4 emission at T = 28 K, and n = 9.0$\times$10$^3$ cm$^{-3}$. It also provides half of the [C~{\sc ii}] emission at T = 530 K, n = 260 cm$^{-3}$. At depths where these lines arise, including the high-J CO lines, grain photoelectric heating dominates and we find no need for additional and exotic heating processes. For S~106 we consider only the (a) model which neglects the extra, high-J component. For  CO 13-12 we find T = 214 K, n = 1.4$\times$10$^6$ cm$^{-3}$, and [O~{\sc i}] 63 \micron\ we find T = 2070 K, n = 7.9$\times$10$^4$ cm$^{-3}$. For the CO 6-5 emission and [C~{\sc ii}] we find T = 40 K, n = 2.7$\times$10$^4$ cm$^{-3}$ and T = 370 K, n = 1.2$\times$10$^3$ cm$^{-3}$ respectively. Again, we find that photoelectric heating dominates in regions where these are formed.

The high density component dominates the overall CO ladder fluxes for upper levels greater than J$_u$ $\simeq$ 9, and therefore the filling factor of the high density component is well constrained by our fits at $\sim$ 0.62 and $\sim$ 0.57 for S~106 and IRAS~23133+6050 respectively. For the low density components factors of 4.6 and 5.1 were found for S~106 and IRAS~23133+6050 respectively, strongly suggesting that additional low density PDRs are present along the line of sight. Finally, the filling factor of the extra very high density PDR component included to fit the highest J S~106 lines was 0.006. The parameters of the various components which make up the composite models are given in Table~\ref{tables:pdrmodprops}.

\begin{table*}
\caption{\label{tables:pdrmodprops} Parameters of best-fitting numerical PDR models}
\begin{center}
\begin{tabular}{l c c c c c}
\hline\hline
Model & Component  & Type$^a$ & Thermal Pressure or Density & \Gnaught & Filling Factor \\
      &            &          & [K cm$^{-3}$] or [cm$^{-3}$]           &          & \\
\hline
\\
IRAS 23133+6050 & 1 & T & 1.5(5) & 2.0(4) & 5.1 \\
IRAS 23133+6050 & 2 & T & 1.7(8) & 1.6(5) & 0.57 \\
\\
S~106 (a) & 1 & T & 5.0(5) & 1.6(4) & 4.6 \\
S~106 (a) & 2 & T & 1.8(8) & 1.6(5) & 0.62 \\
\\
S~106 (b) & 1 & T & 5.0(5) & 1.6(4) & 4.6 \\
S~106 (b) & 2 & T & 1.8(8) & 1.6(5) & 0.62 \\
S~106 (b) & 3 & D & 1.0(8) & 1.6(5) & 0.006 \\
\hline
\end{tabular}

\bigskip
$^a$: Constant thermal pressure (T) or constant density (D).
\end{center}
\end{table*}

In addition to the CO line fluxes, the models also report the fluxes of each of the fine-structure lines in the SPIRE/PACS range. In Table~\ref{tables:modflux} we show the ratio of the measured line fluxes and those predicted by the numerical models. For most lines the model does a good job of replicating the observed line flux within a factor of 2-3. As the [O~{\sc i}] 63 \micron\ and [C~{\sc ii}] 157 \micron\ lines are commonly employed as tracers of shocks, we investigated whether the underprediction of the fluxes of these lines could be due to a contribution of a possible shock. We have estimated the possible contribution to these lines from shocks in S~106 by measuring the fraction of the total $^{12}$CO flux present in the high excitation component which possibly represents a shock compared to the dominant, cooler, components which are well fit by PDR models. This ratio is around 5\%, showing that even in S~106 where there is evidence for a shock in the CO lines, it is over an order of magnitude smaller in flux than the PDR emission in the CO lines. Assuming that the effect on the ionic lines would be of similar magnitude, shocks cannot explain the discrepancy between our models and observations. In fact, the differences in line flux estimates between different PDR models are of the same order of magnitude as the discrepancies we observe between our numerical model fluxes and the observed line fluxes: therefore we conclude that the extra emission in these lines is unlikely to be due to shocks. In particular the IRAS~23133+6050 [C~{\sc ii}] 157 \micron\ line provided the most discrepant line flux ratio of around four, which is somewhat strange given both sources were fit with similar parameters and only the [C~{\sc ii}] in IRAS 23133+6050 shows a large deviation from the model. We suspect that additional low density [C~{\sc ii}] line emission lies along the line of sight to the source, or alternatively this could reflect complexity in the geometry of the source which could perhaps be resolved by higher spectral resolution measurements which would reveal the velocity structure.

\begin{table}
\caption{\label{tables:modflux} Observed / Model Line Flux Ratios.}
\begin{center}
\begin{tabular}{l c c c c}
\hline\hline
Line & Wavelength  & IRAS & S~106 (a)  & S~106 (b) \\
     & [\micron]   & 23133+6050\\
\hline
\\
   ~[C~{\sc i}] & 609.14 &   0.55  &  0.43 & 0.43 \\
   ~[C~{\sc i}] & 370.42 &   0.35  &  0.29 & 0.29 \\
  ~[C~{\sc ii}] & 157.75 &   4.04  &  1.65 & 1.64 \\
   ~[O~{\sc i}] & 145.54 &   2.43  &  1.89 & 1.84 \\
   ~[O~{\sc i}] &  63.18 &   0.88  &  0.71 & 0.71 \\
\hline
\end{tabular}
\end{center}
\end{table}



\section{Discussion}\label{sec:disc}

\subsection{Comparison with Galactic PDRs}\label{sec:comp}

\begin{sidewaystable*}
\vspace{17cm}
\caption{\label{tables:galpdrs} Summary of published properties of Galactic PDRs sorted by the spectral type of the primary exciting star.}
\begin{center}
\begin{tabular}{lc c c c c c c c p{2.5cm}}
\hline\hline
       & Spectral	 & \multicolumn{2}{c}{Inter-clump Phase}& \multicolumn{2}{c}{Dense Phase}& \\
Object & Type	 	 & \Gnaught		& $n$		&  \Gnaught$^a$       & $n$          & $\epsilon$ & I$_{158}$ / I$_{63}$$^b$ & I$_{145}$ / I$_{63}$$^b$ & References \\
       & 	         & [Habing units]	& [cm$^{-3}$]	&  [Habing units] & [cm$^{-3}$]  & $\times$ 10$^{-3}$  \\
\hline
\textit{\HII\ Region PDRs:}\\
30 Dor                  & O3$^c$   & 7.8 $\times$ 10$^3$   & 6.8 $\times$ 10$^3$       & --                    & --                   & --        & --   & --   & 1, 2\\
M17			& O4$^c$   & 5.6 $\times$ 10$^5$   & 3.0 $\times$ 10$^4$       & --                    & 5.0 $\times$ 10$^5$  & 2.3       & 0.17 & 0.13 & 3 \\
M17			& O4$^c$   & --                    & 1.0 $\times$ 10$^3$       & --                    &  --                  & --        & --   & --   & 4 \\
M17    			& O4$^c$   & --                    & --                        & 1--8 $\times$ 10$^4$  & 2.0 $\times$ 10$^7$  & --        & --   & --   & 5\\
W49A                    & O5$^c$   & 3.2 $\times$ 10$^5$   & 1.0 $\times$ 10$^4$       & --                    & --                   & 0.14-0.36 & 0.34 & 0.15 & 6\\
Orion Bar 		& O6       & 4.4 $\times$ 10$^4$   & 2.0 $\times$ 10$^5$       & --                    & $\sim$ 10$^7$        & --        & 0.11 & 0.07 & 7, 8, 9, 10\\
Cepheus B               & OB assoc & --                    & --                        & 1.7 $\times$ 10$^3$   & 1.0 $\times$ 10$^6$  & --        & --   & --   & 11\\
S~106$^d$               & O7-9     & 1.6 $\times$ 10$^4$   & 0.6 -- 3 $\times$ 10$^4$  & 1.6 $\times$ 10$^5$   & 3.1 $\times$ 10$^5$  & 1.0       & 0.13 & 0.15 & This paper\\
S~106                   & O8       & --                    & --                        & > 10$^5$              & 10$^5$ -- 10$^6$     & --        & --   & --   & 12\\          
S~106			& O7-9     & > 10$^{3}$            & $\sim$3 $\times$ 10$^4$   & 1--2 $\times$ 10$^5$  & 1--3 $\times$ 10$^5$ & --        & 0.11 & 0.17 & 13\\
Sgr B2                  & O7       & 5 $\times$ 10$^3$     & 5 $\times$ 10$^3$         & --                    & --                   & --        & --   & --   & 14\\
IRAS~23133+6050$^d$     & O8.5-9.5 & 1.0 $\times$ 10$^4$   & 0.3--1.2 $\times$ 10$^4$  & 1.6  $\times$ 10$^5$  & 3.1 $\times$ 10$^5$  & 1.1       & 0.21 & 0.15 & This paper\\
NGC 2024 - IRS 2b       & O8-B2    &$\sim$5 $\times$ 10$^4$& $\sim$5 $\times$ 10$^5$   & --                    & --                   & --        & --   & --   & 15, 16\\ 
S~140			& B0       & 255                   & 2.0 $\times$ 10$^3$       & --                    & 2.0 $\times$ 10$^4$  & --        & --   & --   & 17\\   
W3 IRS 5                & B0.5     & 1.0 $\times$ 10$^6$   & 1.5 $\times$ 10$^4$       & --                    & 1.8 $\times$ 10$^5$  & --        & --   & --   & 18, 19\\
S~125                   & B0.5     & 9.0 $\times$ 10$^2$   & 1.8 $\times$ 10$^2$       & --                    & --                   & --        & --   & --   & 20\\
\\
\textit{Reflection Nebula PDRs:}\\
NGC 2023 	        & B1.5     & 1.5 $\times$ 10$^4$   & 2.0 $\times$ 10$^4$       & --                    & --                   & 2.6       & 0.19 & 0.06 & 21\\
NGC 2023	        & B1.5     & --                    & --                        & 1.7 $\times$ 10$^4$   &  2.0 $\times$ 10$^5$ & --        & --   & --   & 22\\
NGC 2316	        & B2-3     & 8.1 $\times$ 10$^4$   & 1.0 $\times$ 10$^4$       & --                    & --                   & 1.8       & 0.42 & 0.13 & 23\\ 
NGC 7023	        & B3       & 2.6 $\times$ 10$^3$   & 4.0 $\times$ 10$^3$       & --                    & --                   & 2.9       & 0.85 & --   & 24\\
NGC 7023	        & B3       & 2.0 $\times$ 10$^3$   & 7.0 $\times$ 10$^3$       & --                    & --                   & --        & --   & --   & 25\\
NGC 2245	        & B7-8     & 3.5 $\times$ 10$^2$   & 5.0 $\times$ 10$^3$       & --                    & --                   & 7.6       & 0.75 & 0.14 & 23\\
\hline
\end{tabular}
\end{center}
\smallskip
$^a$: If different from the interclump medium radiation field.\\
$^b$: Observed line ratios, typically interpreted as density diagnostics.\\ 
$^c$: For these sources the PDR is illuminated by a cluster of stars, in the spectral type column we have indicated the hottest star detected in the literature (e.g. \citealt{2008ApJ...686..310H} for M17 and \citealt{1997ApJ...482..307D} for W49A).\\
$^d$: For S~106 and IRAS~23133+6050 we quote the Interclump \Gnaught\ and $n$ values indicated using the combinations of [O~{\sc i}] 63, 145 \micron, [C~{\sc ii}] 158 \micron\ and the total far-infrared flux reported in rows 3-4 of Table~\ref{tables:diags}. The dense phase (clump) values arising from the average of the high-J CO lines and the cooling lines / FIR flux (rows 7-8 in Table ~\ref{tables:diags}).\\
\tablebib{(1) \citet{1999AAS..137...21B}; (2) \citet{1995ApJ...454..293P}; (3) \citet{1992ApJ...390..499M}; (4) \citet{2010AA...510A..87P}; (5) \citet{2013ApJ...774L..14S}; (6) \citet{2001AA...376.1064V}; (7) \citet{1985ApJ...291..747T} ; (8) \citet{1993Sci...262...86T}; (9) \citet{1997ApJ...481..343H}; (10) \citet{1982ASSL...90.....G}; (11) \citet{2012AA...542L..17M}; (12) \citet{2000AA...358.1035V}; (13) \citet{2003AA...406..915S}; (14) \citet{2004ApJ...600..214G}; (15) \citet{2003AA...404..249B}; (16) \citet{2000AA...358..310G}; (17) \citet{1997AA...323..953S}; (18) \citet{2004AA...424..887K};(19) \citet{2005ApJ...622L.141M}; (20) \citet{2003AA...406..155A}; (21) \citet{1997ApJ...478..261S}; (22) \citet{2011ApJ...741...45S}; (23) \citet{2002ApJ...578..885Y}; (24) \citet{1988ApJ...334..803C}; (25) \citet{2010AA...521L..25J}. }

\end{sidewaystable*}

While the Orion region remains the prototypical example of a PDR, many other examples have been observed and interpreted using PDR models. In Table~\ref{tables:galpdrs} we show the properties of the regions studied here alongside a sample of directly comparable PDRs associated with \HII\ regions, as well as the properties of some PDRs generated by cooler stars\footnotemark. If we consider initially the properties of the interclump phase, our sources lie somewhere between these two classes of sources, with lower overall radiation fields than M17, W49A or Orion -- as one might expect given their relatively cooler exciting stars and the fact that M17 and W49A are illuminated by clusters of O stars. In fact, the central clusters of both M17 and W49A both possess several examples of stars with the spectral type listed in Table~\ref{tables:galpdrs}. The \Gnaught\ values we find for our sources are actually much closer to those found for the reflection nebulae, particularly that of NGC 2023. In terms of density, the interclump medium for our PDRs is consistent with the other examples shown in Table~\ref{tables:galpdrs}, with the exception of the Orion bar which presents a density around an order of magnitude higher. For the dense clump component the density derived for S~106 is very similar to that found before for S~106 by \citet{2003AA...406..915S}, and also close to the value quoted by \citet{1992ApJ...390..499M} for M17. These values are typical of those quoted for the clumps in a dense PDR, however the value of $\sim$ 10$^7$ cm$^{-3}$ for the dense component of the Orion bar PDR by \citet{1993Sci...262...86T} is actually very similar to the very high density PDR component we invoked in our numerical models to attempt to fit the highest-J CO lines in S~106. This would suggest that the high-J component might also be present for the Orion bar region. Hints of this phenomenon may be visible in the SPIRE observations of the Orion bar presented by \citet{2010A&A...518L.116H}, where the presented \verb+RADEX+ fits to the $^{12}$CO lines begin to deviate in the mid-J regime (J $\sim$ 12), however the region was not observed using the PACS spectroscopy range scan mode employed for our objects, so observations with SOFIA or other FIR observatories would be required to confirm this hypothesis.

\footnotetext{The optical part of such systems are usually referred to as reflection nebulae, as the primary optical emission from such nebulae consists of reflected starlight rather than emission lines as one would expect from \HII\ regions.}

\begin{figure}
	\begin{center}
	\includegraphics[width=9cm]{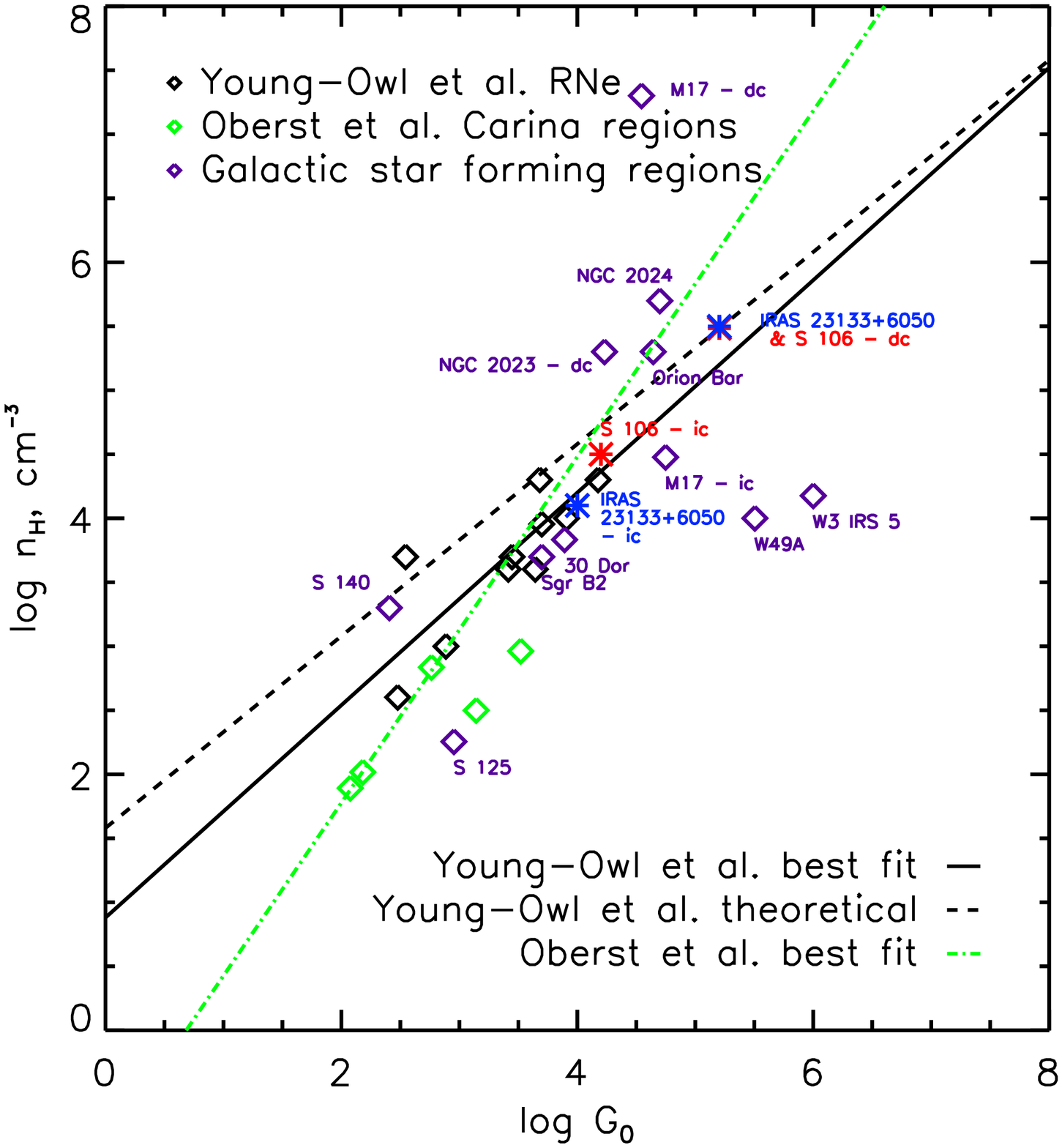}
	\end{center}
	\caption{Correlations between \Gnaught\ and density $n_0$ found by \citet{2002ApJ...578..885Y} and \citet{2011ApJ...739..100O}. Reflection nebula points from \citet{2002ApJ...578..885Y},  Galactic PDRs as stated in Table~\ref{tables:galpdrs}. IRAS~23133+6050 and S~106 fit into the trend between the larger \HII\ regions with hotter exciting stars and the RNe PDRs with B type exciting stars. Dashed and solid black lines: theoretical prediction for the relationship between radiation field and density for a pressure bound \HII\ region and the best fit to the \citet{2002ApJ...578..885Y} data. Green dot-dashed line: best fit to the Carina nebula data of \citet{2011ApJ...739..100O}. For a selection of sources we have also included the points representing the dense PDR component (denoted `dc') as opposed to the default values representing the interclump medium (denoted `ic').}
	\label{fig:yo_corr}
\end{figure}

Several groups have shown that there is a relationship between the incident radiation field and the PDR density for a variety of objects. Studies have examined this relationship for a sample of reflection nebula PDRs \citep{2002ApJ...578..885Y}, while a more recent study found a slightly different relationship for spatially resolved regions of the Carina nebula \citep{2011ApJ...739..100O}. If we consider the upper bound values for the inter-clump density in Table~\ref{tables:galpdrs}, our two sources fall almost perfectly on both of these trends, as is shown in Figure~\ref{fig:yo_corr}. The dashed line represents the expected conditions for systems in a state of equilibrium between the pressure and temperature of the \HII\ region and PDR respectively under the assumption that the \HII\ region is a Stromgren sphere and has an electron temperature of 8000 K (\citealt{2002ApJ...578..885Y}; referred to as `Young-Owl et al. theoretical' in Figure~\ref{fig:yo_corr}). Conversely, \citet{2011ApJ...739..100O} suggest that their best fit line represents systems which are density bounded as opposed to the ionization bounded RNe, and hence they found a different relationship. We have also included the \HII\ region PDR sources from Table~\ref{tables:galpdrs} while the RNe sources in Table~\ref{tables:galpdrs} form part of the \citet{2002ApJ...578..885Y} sample and are represented by the black diamonds in Figure~\ref{fig:yo_corr}. Our sources are consistent with the \HII\ region PDRs with late O type exciting stars and the RNe formed by early B stars.  Amongst the \HII\ region PDRs with hotter exciting stars there seems to be a greater spread in properties away from the trends, with W49A and W3 IRS 5 in particular having either higher \Gnaught\ or lower $n_H$ by at least an order of magnitude than predicted by either relationship. We have also included the properties of the dense phase for several sources, denoted as `dc' in Figure~\ref{fig:yo_corr}. For our sources and NGC 2023, the properties of the dense phases fall very near either of the relationships discussed, the dense component of M17 however falls well above even the matter bounded trend given by \citet{2011ApJ...739..100O} by several orders of magnitude, this may be due to the dense clumps being independently gravitationally bound or transient objects -- that would therefore have much higher densities than would be expected for pure pressure equilibrium between the clumps and their surroundings. Overall it seems that the simpler sources with a dominant exciting star (e.g. IRAS~23133+6050, S~106, NGC~2023, S~125, S~140) agree well with the trends, while the much more complex sources (e.g. W49A, M17, W3) have much more variance. However one of the most complex sources, 30 Dor, is consistent with both predictions, so the possible trend between simple/complex sources could easily be a manifestation of small number statistics.

It is expected that the flux ratio of the 158 \micron\ [C~{\sc ii}] and the 63 \micron\ [O~{\sc i}] lines should decrease with increasing \Gnaught\ and $n$ as the [O~{\sc i}] line becomes the dominant source of cooling. We see this progression between our two sources, with IRAS~23133+6050 having a cooler ionizing star and lower \Gnaught\ and a correspondingly higher 
 $^{\textrm{[C~{\sc ii}]}}/_{\textrm{[O~{\sc i}]}}$ ratio than S~106 by around a factor of two. The two values we find for this ratio fit nicely into the progression seen in Table~\ref{tables:galpdrs}, with lower values for the \HII\ regions, (with the puzzling exception of W49A although this may be due to the large physical size of W49A in comparison to the other objects) and higher values for the RNe PDRs. As with the comparisons between the radiation field and densities of the regions, the closest Galactic analog of IRAS~21333+6050 we measured is actually the reflection nebula NGC 2023. This likely arises because the O8.5-9.5 star illuminating IRAS~21333+6050 has a surface temperature relatively close to that of the B1.5 central star of NGC 2023. On the other hand, S~106 is closer to the \HII\ region PDRs listed, and actually shows the same line ratio as the Orion bar region.

In context of other Galactic PDRs then, on the basis of the spectral types of the exciting stars alone the PDRs studied here might be expected to have properties between those of the massive star forming regions and clusters and the RNe PDRs, and this is indeed what we find. With both regions having properties similar to the PDRs generated by the hottest B stars, especially IRAS~23133+6050, which is very similar to the PDR of the RNe NGC 2023. The general consistency between our two objects can also be explained using similar logic. Both sources have similar \HII\ region electron densities and sizes, and similar types of exciting stars, around O8-9, although from the literature presented in section~\ref{sec:targs} it seems likely that the star illuminating S~106 is slightly hotter by about one subtype. Despite this, both sources have very different morphologies, with S~106 having a very complex dual lobed \HII\ region and a small stellar disk region surrounded by a large extended PDR, while IRAS~23133+6050 is still an ultra-compact \HII\ region yet to break out of its natal molecular cloud. This dichotomy poses an interesting problem and might constrain models for the evolution and dynamics of \HII\ regions and PDRs.


\subsection{S~106}

S~106 in particular has a rich vein of literature and prior studies, of which a few discuss its PDR specifically (for example \citet{2000AA...358.1035V} and \citet{2003AA...406..915S} listed in Table~\ref{tables:galpdrs}). We discussed these studies in terms of comparing the measured surface brightnesses for the various PDR cooling lines in Section~\ref{sec:line_flux}, here we discuss the overall results of those studies and how they compare to our own. In general there is broad agreement between the literature findings and our results for S~106, in that there is a clumpy PDR consisting of dense clumps irradiated by a strong UV field producing the high-J CO lines, along with a less dense PDR phase permeating the region. We have extended these findings by presenting the full CO ladder available to the PACS and SPIRE Herschel instruments, which have revealed the presence of a further, even higher density, component. The key difference between the prior studies and our results here is that we are probing only the innermost region of S~106 (a circular aperture of radius 21.5\arcsec), whereas \citet{2000AA...358.1035V} used ISO-LWS spectra of the central region with a beam some three to five times larger than our Herschel observations. The \citet{2000AA...358.1035V} concluded that the S~106 PDR was of very high density and experienced an extremely intense UV radiation field (\Gnaught > 10$^5$) despite measuring a similar set of the common PDR cooling lines as we presented earlier from the Herschel observations and found evidence of a more moderate environment. The key difference arises because \citet{2000AA...358.1035V} also considered ISO-SWS observations including the [Si~{\sc ii}] 34.8 \micron\ line, the strength of which could only be explained by invoking stronger radiation fields and higher densities. 

A subsequent study by \citet{2003AA...406..915S} spatially resolved the larger scale structure of S~106 PDR cooling line emission with pixel scales of 20-40\arcsec using the FIFI instrument on the KAO, and also including ISO-LWS spectra. \citet{2003AA...406..915S} found density variations within the PDR in a similar manner to those described in this work, and very similar parameters for the two phases were recovered. The only slight difference arises in the density of the dense component, in which we found a value slightly above the range quoted by \citet{2003AA...406..915S}. We also find a value for the average UV radiation field experienced by the interclump medium of \Gnaught\ = 1.6 $\times$ 10$^4$ while \citet{2003AA...406..915S} give a lower limit of \Gnaught\ > 10$^3$. The fact that we recover very similar values for the density and incident radiation field for the interclump medium in the region immediately adjacent to the star as was found by previous studies as averages for the entire region can then be easily explained by inspection of the maps presented by \citet{2003AA...406..915S} which show that the emission from S~106 is very strongly peaked near the exciting star in the center of the region. By using the full Herschel CO ladder we have extended this picture to include a further excitation component, which may represent a further, yet higher density, PDR component or a shock -- a conundrum we will discuss in the following subsection.  


\subsection{The diagnostic value of the CO ladder}\label{sec:discCO}

We have used the observed CO fluxes in order to constrain the PDR properties of our sources. The shape of the CO ladder (as well as the curves in the rotation diagrams) provides the number of components with distinct physical conditions present in each source. In combination with the simple diagnostics which use the standard PDR lines, the ladder was then used to find the temperature of each component using \verb+RADEX+ models. Subsequently we used the CO fluxes to guide the application of numerical PDR models -- leading to a reasonable fit to all of the other observables. However, there are several complexities to this picture, firstly: to what extent do the results of the rotation diagram analysis make sense in terms of our more complex models? It is not clear that the level populations are in LTE while the lower J lines may be optically thick and the temperature distribution is not constant, so it might be expected that the results we find for the rotation diagrams compare poorly to the more detailed approaches. Secondly: we have detected the presence of an extra excitation component in the high-J S~106 CO lines which could be ascribed to either very-high density/thermal pressure PDR, or a strong shock.

The rotation diagram fits to both sources revealed very similar rotation temperatures of around 190 K (188 and 190 K for IRAS~23133+6050 and S~106 respectively) for the mid-J $^{12}$CO lines. In Section~\ref{sec:radex_rot}, we compared these temperatures to those found using \verb+RADEX+ fits to the respective $^{12}$CO ladders. Using a density of 10$^{5.6}$ cm$^{-3}$ as indicated by the diagnostic diagrams gave temperatures of around 100 K warmer. We also found reasonable fits to the CO ladders using \verb+RADEX+ with higher densities and a temperature of 200 K. In light of the results of our numerical models, these results are actually reasonable estimates. For example, for IRAS~23133+6050 we found that the peak CO line emission emanated from a region of T = 217 K at a density roughly three times higher than that predicted by the simple diagnostic diagrams. This concurrency suggests that the skepticism with which we referred to the rotation diagram figures was perhaps unwarranted; in fact, for these data one could use the rotational temperature found using the rotation diagrams as a guide for using \verb+RADEX+ models to find the PDR density in the vicinity of the CO molecules.

In a more general sense the CO ladders we find are typical, and are routinely found in extragalactic contexts. This is because any observation of a star-forming galaxy  will inevitably include star forming complexes like S~106 and IRAS~23133+6050. However, such observations may well include regions dominated by other physical processes, including analogues of PDRs dominated instead by X-rays or cosmic rays (XDRs or CDRs), PDRs with enhanced mechanical heating (mPDRs) or an environment dominated by shocks (e.g., \citealt{2010A&A...518L..42V, 2011ApJ...743...94R, 2014A&A...568A..90R} and references therein). Each of these processes has some effect on both the relative strengths of the CO ladder and the fine-structure lines. The challenge in interpreting such observations then is to tease apart the different effects and attempt to create a coherent model of the environment. In contrast our sources are relatively simple and do not seem to be dominated by the wide variety of other physical processes that can be important for star-forming galaxies or AGN. Therefore, the extreme conditions we infer in our sample represent a baseline for the more complex sources, in that such conditions would be expected in the PDRs associated with massive star formation in these extragalactic sources.

There appear to be two plausible explanations for the extra heating in the very high-J (J $>$ 20) CO lines seen in S~106; shocks or a very high density PDR. Previous work on S~106 with respect to shocks is somewhat conflicting. \citet{2000AA...358.1035V} found that the IR emission from S~106 was dominated by PDR emission on the basis of the  [S~{\sc i}] 25.2 \micron\ / [Si~{\sc ii}] 34.8 \micron\ line ratio, which is usually thought to be representative of shocks. However, the \citet{2000AA...358.1035V} study used ISO-SWS observations with a comparatively large beam - possibly implying that the environment is dominated by PDR material, rather than exclusively PDR material. A later spatially resolved study by \citet{2005AA...436..569N} found evidence of thermalization in near-IR H$_2$ line ratios, which could be evidence of shocks in the S~106 region, which would be included in the beam of our observations. Such ratios could also be produced by a very high density PDR (e.g., \citealt{1990ApJ...352..625B}), although \citet{2005AA...436..569N} conclude that this cannot be the case for one of the thermalized regions, as it lacks detections of Br$\gamma$, PAHs and continuum emission. As we noted earlier, shocks are common in star formation regions of most masses (e.g., \citealt{2013A&A...557A..22S, 2014A&A...562A..45K}). The bipolar morphology of S~106 lends further credence to the idea that shocks are present in S~106 and are contributing to the extra excitation observed in the CO ladder.

In contrast, the evidence supporting the alternative view -- that the extra excitation arises from the high density PDR itself -- is somewhat limited and circumstantial. If taken at face value, the center panel of Fig.~\ref{fig:modelfit} and the negligible differences in terms of the fine-structure line fluxes between the two S~106 models in might appear to support the idea that the emission arises from a very high density PDR. However, we found that a good fit to the high J lines could be only reached using an additional PDR model with a density of 10$^8$ cm$^{-3}$ and \Gnaught\ of 1.6 $\times$ 10$^5$. Such densities could perhaps be plausible in two circumstances. Firstly, the small circumstellar disk around IRS4 in the core of S~106; secondly, small dense clumps interspersed within the main PDR. However, both of these scenarios could be problematic. The first possibility would require a much higher values of \Gnaught\ than used in our model by virtue of being much closer to the primary star, however the second circumstance would require small clumps of PDR material with densities several orders of magnitude beyond any previously observed. Clearly further studies will be required to decisively settle this issue. Nevertheless, the general shape of the CO ladder fluxes is very similar to Fig.~5 of \citet{1990ApJ...365..620B}, in which at high densities and radiation fields, the high-J CO lines can be excited in a thin layer on the surface of PDRs at temperatures of several hundred K. The transition between the two regimes is not at precisely the same J$_U$, however the shape of the curve is almost identical.



\section{Conclusions and Summary}\label{sec:concsum}

We have obtained full Herschel -PACS and -SPIRE spectra of the star forming regions S~106 and IRAS~23133+6050 and analyzed them in terms of the emission from their PDRs. Simple PDR diagnostic diagrams were employed to find the physical conditions and hints of the physical structure of the two PDRs. These results were tested using a detailed numerical model of PDR physics. The study reveals that the PDRs around S~106 and IRAS~23133+6050 are remarkably similar in some respects, despite being dramatically different types of objects -- S~106 is a large, complex star formation region, with many components, while IRAS~23133+6050 is a relatively simple cometary ultra-compact \HII\ region. Despite these differences, both objects were found to have a similar range of physical conditions and structures, in common with many other PDR observations and models. The structure of each region, implied by our observations, is that of a classical PDR with intermediate properties (\Gnaught=10$^{4}$, $n$=10$^{4}$ cm$^{-3}$) and with embedded clumps of much denser material (\Gnaught=10$^{5}$, $n$=10$^{6}$ cm$^{-3}$). We employed numerical models which were fitted to the $^{12}$CO ladder fluxes, initially to determine the temperature of the PDR environments from the peak of the CO ladder, as well as to investigate whether the conditions indicated by the simple diagnostics were consistent with those indicated by a full numerical treatment. During this process it was noticed that the S~106 CO ladder possesses an extra excitation component at the high J lines. This component is only reconcilable with PDR models upon invoking extreme densities, so we tentatively ascribe shocks as a probable cause. Ultimately though, both PDRs studied possess very similar properties to those of the Orion complex, which is surprising as Orion and the two objects studied here are in very different evolutionary stages. The youngest object, IRAS~23133+6050, is still embedded in its natal cloud, while in Orion the cluster of exciting stars have emerged from this phase and are steadily eroding their environment. The consistency of properties on various PDRs between objects and in time underlines their importance as a key phase of the ISM.

\begin{acknowledgements} 
DJS thanks the referee, N. Schneider, for a very thorough and thoughtful referee report which has greatly improved the discussion sections of the paper. DJS also thanks E. Polehampton at Rutherford Appleton Laboratories for his assistance in reducing the SPIRE data and understanding the Herschel SPIRE beam profiles.  

DJS and EP acknowledge support from an NSERC Discovery Grant and an NSERC Discovery Accelerator Grant. 
M.G.W. was supported in part by NSF grant AST-1411827.

Studies of interstellar chemistry at Leiden Observatory are supported through advanced-
ERC grant 246976 from the European Research Council, through a grant by the Dutch
Science Agency, NWO, as part of the Dutch Astrochemistry Network, and through the
Spinoza premie from the Dutch Science Agency, NWO.

C.B. is grateful for an appointment at NASA’s Ames Research Center through San José State University Research Foundation (NNX14AG80A).

J. C. acknowledges support from an NSERC Discovery Grant.

This research has made use of NASA's Astrophysics Data System Bibliographic Services.

\end{acknowledgements}

\bibliographystyle{aa}

\end{document}